\documentclass{aa}
\usepackage{times}
\usepackage{latexsym}
\usepackage{epsfig}

\def\xip{$\xi$~Per}
\def\lab{$\lambda$~Cep}
\def\cam{$\alpha$~Cam}
\def\lor{$\lambda$~Ori}
\def\zor{$\zeta$~Ori}
\def\cyg{68~Cyg}
\def\mon{15~Mon}
\def\sao{HD~34656}
\def\cep{19~Cep}
\def\lac{10~Lac}

\def\heii{He~{\sc ii}}
\def\civ{C~{\sc iv}}

\def\niv{N~{\sc iv}}
\def\nv{N~{\sc v}}

\def\siiv{Si~{\sc iv}}
\def\kms{km~s$^{-1}$}
\def\0{\hspace*{0.5em}}
\def\vinf{$v_{\infty}$}

\def\ha{H$\alpha$}

\begin{document}

\thesaurus{07(08.05.1; 08.13.1; 08.13.2; 08.15.1; 13.21.5)}

\title{Long- and short-term variability in O-star winds\thanks{Based
on observations by the International Ultraviolet Explorer,
collected at NASA Goddard Space Flight Center and Villafranca
Satellite Tracking Station of the European Space Agency}}
\subtitle{II. Quantitative analysis of DAC behaviour}
\author{L.~Kaper\inst{1,2}
   \and H.F.~Henrichs\inst{2}
   \and J.S.~Nichols\inst{3}
   \and J.H.~Telting\inst{4,2}}

\offprints{L. Kaper.  E-mail: lexk@astro.uva.nl}

\institute{European Southern Observatory, Karl Schwarzschild Str.\ 2, 
           D-85748 Garching bei M\"{u}nchen, Germany
   \and    Astronomical Institute ``Anton Pannekoek", University of Amsterdam,
           Kruislaan 403, 1098 SJ Amsterdam, Netherlands
   \and    Harvard-Smithsonian Center for 
           Astrophysics, 60 Garden Street, Cambridge MA 02138, U.S.A.
   \and    Isaac Newton Group of Telescopes, NFRA, Apartado 321, 38700 
           Santa Cruz de La Palma, Spain}

\date{Received: July 30, 1998; Accepted: December 21, 1998}

\maketitle

\begin{abstract}
A quantitative analysis of time series of ultraviolet spectra from a
sample of 10 bright O-type stars (cf.\ Kaper et al.\ \cite{KH96},
Paper~I) is presented. Migrating discrete absorption components
(DACs), responsible for the observed variability in the UV resonance
doublets, are modeled. To isolate the DACs from the underlying P~Cygni
lines, a method is developed to construct a template
(``least-absorption'') spectrum for each star. The central velocity,
central optical depth, width, and column density of each pair of DACs
is measured and studied as a function of time.

It turns out that the column density of a DAC first increases and
subsequently decreases with time when the component is approaching its
asymptotic velocity. Sometimes a DAC vanishes before this velocity is
reached. In some cases the asymptotic DAC velocity systematically
differs from event to event.

In order to determine the characteristic timescale(s) of DAC
variability, Fourier (CLEAN) analyses have been performed on the time
series. The recurrence timescale of DACs is derived for
most targets, and consistent results are obtained for different
spectral lines. The DAC recurrence timescale is interpreted as an
integer fraction of the stellar rotation period. In some datasets the
variability in the blue edge of the P~Cygni lines exhibits a longer
period than the DAC variability. This might be related to the
systematic difference in asymptotic velocity of successive DACs. 

The phase information provided by the Fourier analysis confirms the
expected change in phase with increasing velocity. This supports the
interpretation that the DACs are responsible for the detected
periodicity. The phase diagram for the O~giant \xip\ shows clear
evidence for so-called ``phase bowing'', which is an observational
indication for the presence of curved wind structures like corotating
interaction regions in the stellar wind. An important difference with
the results obtained for the B~supergiant HD 64760 (Fullerton et
al. \cite{FM97}) is that in this O~star the phase bowing can be
associated with the DACs. No other O~stars in our sample convincingly
show phase bowing, but this could be simply due to the absence of
periodic signal and hence coherent phase behaviour at low wind
velocities.


\end{abstract}

\section{Introduction}

Variability is a fundamental property of the radiation-driven winds of
early-type stars. Discrete absorption components (DACs) are the most
prominent features of wind variability. They migrate from red to blue
in UV P~Cygni lines, narrowing in width when they approach the
terminal velocity \vinf\ of the stellar wind. Obviously, DACs cannot 
be observed in saturated P~Cygni profiles; however, the steep blue edges
of these profiles often show regular shifts of up to 10\% in
velocity. Edge variability is probably related to the DAC behaviour,
but the precise phase relation has not been unraveled yet. For a more
extensive introduction and various examples of the DAC phenomenon we
refer to the first paper in this series (Kaper et al. \cite{KH96},
Paper~I). Reviews on this subject were presented by Henrichs
(\cite{He84},\cite{He88}), Howarth (\cite{Ho92}), Kaper \& Henrichs (\cite{KH94}), Prinja (\cite{Pr98}), and Kaper (\cite{Ka98}).

\begin{table*}
\caption[]{Main observational properties of the program stars. Notes:
(a) Walborn \cite{Wa72}, except HD~210839 Walborn \cite{Wa73}; (b)
Penny \cite{Pe96}; (c) Prinja et al. \cite{PB90}. A typical
uncertainty in \vinf\ is 50 \kms. For comparison, also the (maximum)
asymptotic velocity of the DACs, $v_{\mbox{\scriptsize asymp}}$ is
given based on our datasets. In the last column we list the number of
highest flux points selected from the total set of spectra to
construct the template spectrum.}
\begin{flushleft}
\begin{tabular}{rllccccc}
\noalign{\smallskip}
\hline
\noalign{\smallskip}
HD & Name & Spectral type$^{\,a}$ & $v \sin{i} ^{\,b}$ & \vinf$^{\,c}$ &
$v_{{\rm asymp}}$ & \#spectra & $n$ \\
 & & & (\kms) & (\kms) & (\kms) & ($N$) & \\ \hline
\noalign{\smallskip}
\hline\noalign{\smallskip}
24912 & $\xi$ Per    & O7.5 III(n)((f)) & 204  & 2330 & 2300 & 124 & 2 \\
30614 & $\alpha$ Cam & O9.5 Ia          & 115  & 1590 &      & \031 & 3 \\
34656 &              & O7 II(f)         & \085  & 2155 & 2125 & \029 & 2 \\
36861 & $\lambda$ Ori A & O8 III((f))   & \066 & 2125 & 2000 & \027 & 3 \\
37742 & $\zeta$ Ori A & O9.7 Ib         & 123  & 1860 & (2100) & \026 & 3 \\
47839 & 15 Mon       & O7 V((f))        & \062 & 2055 & 2280 & \020 & 2 \\
203064 & 68 Cyg      & O7.5 III:n((f))  & 295 & 2340 & 2500 & 149 & 5 \\
209975 & 19 Cep      & O9.5 Ib          & \090 & 2010 & 2080 & \083 & 5 \\
210839 & $\lambda$ Cep & O6 I(n)fp      & 214  & 2300 & (2100) & 123 & 10 \\
214680 & 10 Lac      & O9 V             & \031 & 1120 & \0990 & \023 
& 3 \\ \hline
\noalign{\smallskip}
\hline
\end{tabular}
\end{flushleft}
\end{table*}

One of the key problems is to understand why DACs start to develop and
how they evolve.  Since DAC behaviour is different for different
stars, and even shows detailed changes from year to year for a given
star, a detailed quantitative description of the observed variability
over a significant amount of time (years) is essential.  It should be
emphasized that in the planning phase of this project the sampling
times of the targets have been carefully tuned to each individual
star, based on sample spectra collected in earlier pilot studies.  As
a result, in Paper~I, time series of more than 600 high-resolution
ultraviolet spectra of 10 O-type stars were presented.  In this
follow-up paper we analyse the series of spectra obtained with the
{\it International Ultraviolet Explorer} (IUE) in a quantitative
manner and evaluate the behaviour of DACs and edge variability in the
UV P~Cygni lines.  In the following section we describe the data
analysis and present our methods to isolate and model the DACs in the
ultraviolet spectra. Furthermore, a period-search analysis was
performed to derive the characteristic timescales of both DAC and edge
variability. The results of the modeling and the time series analyses
are presented in section 3. We compare the results obtained for
different stars in section 4 and draw general conclusions.  In
the last section we discuss the observed variability and its regular
behaviour in terms of the corotating interaction regions model of
Cranmer \& Owocki (\cite{CO96}).

\section{Data analysis}

High-resolution ($R=$10,000) ultraviolet spectra of the 10 studied
O-type stars (see Table~1 for our sample) were obtained with the Short
Wavelength Prime (SWP) camera on board the IUE satellite. Time series
of the relevant UV P~Cygni lines are presented in Paper~I, together
with a detailed description of the targets and their observational
history. For each of the 10 O~stars, the high-resolution SWP spectra
form a homogeneous dataset resulting from similar observational
constraints and a uniform reduction procedure (cf.\ Paper~I).

\subsection{Construction of template spectrum}

To isolate the migrating DACs from the underlying P~Cygni profiles, a
template (in this case a ``least-absorption'') spectrum is needed that
represents the undisturbed-wind profile to which the P~Cygni profile
apparently returns after the passage of a DAC (e.g.\ Prinja et al.\
\cite{PH87}, Paper~I). Subsequent division of the individual spectra
by this template then results in quotient spectra, which are used to
model the DACs (see next subsection).

Close inspection of the intrinsic variability in the P~Cygni lines
(cf.\ Paper~I) suggests that the observed changes are mainly due to
variations in the wind column in front of the stellar disk, since the
wind variability is only found in the blue-shifted absorption troughs,
and not in the P~Cygni emission peaks. On the one hand, this is what
one would expect in the case of relatively modest variations in the
density and/or velocity structure of the stellar-wind. To illustrate
this, one could imagine a small blob of gas in the stellar wind
absorbing photons at a particular velocity and reemitting them
isotropically. When this blob is in the line of sight, it leaves a
blue-shifted absorption feature, so that we can ``count'' all the
photons that were scattered out of the absorbing column. For a blob
outside the line of sight, only the few photons that are reemitted in
the direction of the observer can be detected. On the other hand, the
absorbing blob has to cover a significant fraction of the stellar disk
in order to give rise to an observable change in the P~Cygni
absorption.

Our basic assumption is that the changes in the P~Cygni profiles are
caused by variable amounts of additional absorption caused by material
in the line of sight. In terms of Sobolev optical depth, a change in
P~Cygni absorption could be due to a (local) change in the wind
density or the velocity gradient. The simplest
construction of a reference template spectrum is to select the highest
flux point (per wavelength bin) from the available spectra.  Such a
procedure applied to non-variable parts of the spectrum would,
however, yield a template that is systematically too high because of
instrumental and photon noise.  Here we describe a method which
automatically corrects for this overestimation.

\begin{figure}[!ht]
\centerline{\psfig{figure=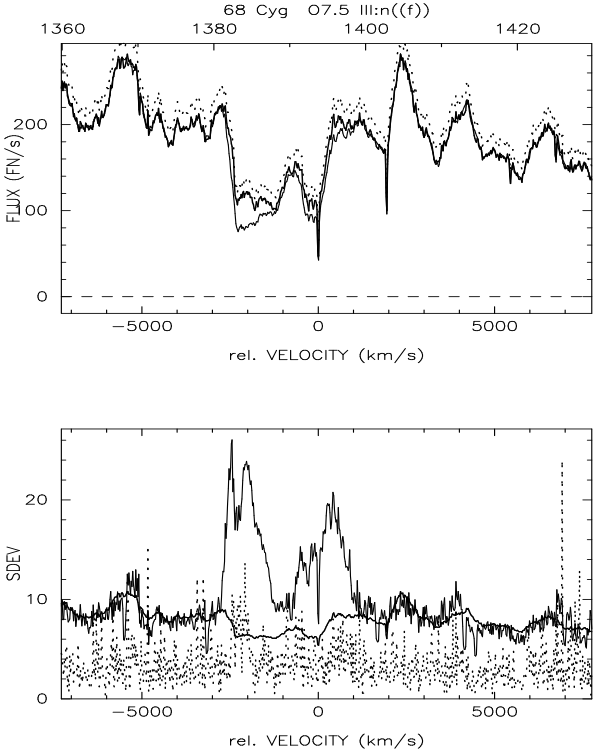,width=8cm,height=10cm}}
\caption[]{To demonstrate our method for the construction of a
``least-absorption'' template, we show the \siiv\ resonance doublet of
the O7.5 giant \cyg. The dotted line indicates the average $\mu$
(upper panel) and standard deviation $\sigma$ (lower panel) of the 5
highest flux points (out of 149 spectra) as a function of
wavelength. The average and standard deviation of the total number of
spectra are represented by a thin line.  The derived template spectrum
($\mu_{0}$, thick line) is identical to the average spectrum in the
continuum, and should represent the undisturbed-wind profile in the
variable \siiv\ line. In the lower panel the difference can be seen
between the instrumental noise (estimated by $\sigma_{0}$, thick line)
and the observed variance (thin line) in the resonance lines due to
wind variability.}
\end{figure}

Our method is based on a paper by Peat \& Pemberton (\cite{PP70})
which documents computer programs for the automatic reduction of
stellar spectrograms, in particular to locate the continuum level in a
spectrogram.  The large number of available spectra ($N$, listed in
Table~1) enables the construction of a ``noise-corrected'' template
spectrum.  We assume that the flux points not disturbed by DACs are
normally distributed with mean value $\mu_{0}(\lambda)$ and variance
$\sigma_{0}^{2}(\lambda)$ within each wavelength bin (taken 0.1~\AA)
at wavelength $\lambda$.  We choose a suitable number $n$ (last column
in Table~1), not too small, but small enough to have the $n$ highest
flux values in every bin sufficiently free from DACs. The underlying
assumption is that if the P~Cygni profiles always return to the same
flux level after passage of a DAC, a sufficient number of flux points
per wavelength bin will be available (provided that the total number
of spectra is large enough) to define the undisturbed-wind profile.

\begin{figure*}[tbp]
\vspace{17cm}
\caption[]{Left: the template spectrum (thick line)
and a ``representative'' spectrum (thin line) including the \siiv\
resonance doublet are shown for the different stars. The units of the
y-axis are in FN/s.  Right: division of the observed spectrum by
the template gives the quotient spectrum (thin line), clearly bringing out 
the DACs in the spectra. The best fit is illustrated as a thick line.}
\end{figure*}

\begin{figure*}[tbp]
\vspace{10cm}
\caption[]{As Fig.\ 2: Template spectrum (thick line) and sample model fits  
(thick line) for the \nv\ resonance doublet.}
\end{figure*}

We can compute $\mu$ and $\sigma^{2}$ of these $n$ highest flux values
for each wavelength bin. We then assume that the ``undisturbed'' flux
points belong to the wing of a normal distribution in any given part
of the spectrum, which allows us to derive $\mu_{0}$ and
$\sigma_{0}^{2}$ from $\mu$ and $\sigma^{2}$.  In parts of the
spectrum that do not vary intrinsically (such as the continuum), the
values obtained for $\mu_{0}$ (and $\sigma_{0}^{2}$) should represent
the average spectrum and the noise; this provides an independent check
of the validity of our assumption.

For our application we used the distribution function
$\phi(x)$  for a normal distribution given by:
\begin{equation}
\phi(x) = \frac{1}{\sqrt{2\pi} \, \sigma_{0}} \, \mbox{e}^{\textstyle -
\left( \frac{x-\mu_{0}} {\sqrt{2} \, \sigma_{0}} \right)^{2}}
\end{equation}
The distribution of the $n$ highest points is a truncated normal
distribution, i.e.\ $x$ running from $w$ to $\infty$, with $w$ from
Eq.\ (5). So we have for the average $\mu$ and variance $\sigma^{2}$
of the $n$ highest flux points:
\begin{equation}
\mu = \frac{\int_{w}^{\infty} x \phi(x) dx}
           {\int_{w}^{\infty}  \phi(x) dx} \: {\mbox{, and}}
\end{equation}
\begin{equation}
\sigma^{2} = \frac{\int_{w}^{\infty} (x-\mu)^{2} \phi(x) dx}
                  {\int_{w}^{\infty}  \phi(x) dx} \, .
\end{equation}
The ratio $k$ is given by:
\begin{equation}
k = \frac{n}{N} = \frac{1}{\sqrt{\pi}} \int_{z}^{\infty} 
    \mbox{\rm e}^{-y^{2}}dy = \frac{1}{2} \mbox{\rm erfc}(z) 
\end{equation}
with
\begin{equation}
z = \frac{w-\mu_{0}}{\sqrt{2} \, \sigma_{0}} \, .
\end{equation}
After substitution of these expressions in Eq.\ (2) we can relate $\mu$ 
to $\mu_{0}$:
\begin{equation}
\mu_{0} = \mu - \frac{\sigma_{0}}{\sqrt{2\pi} \, k} \, 
\mbox{\rm e}^{\textstyle -z^{2}}
\end{equation}
Thus, from $\mu$ and $\sigma_{0}$ we can compute $\mu_{0}$; $z$ is the
inverse complementary error function of $2k$ (Eq.\ 4) and is computed
iteratively, using an adequate approximation for the complementary
error function. Substituting $\mu$ from Eq.\ (6) in Eq.\ (3), we
obtain a relation between $\sigma_{0}^{2}$ and $\sigma^{2}$, the
variance in the $n$ highest points:
\begin{equation}
\sigma_{0}^{2} = \sigma^{2} \left( \frac{2\pi k^{2}}
{2\pi k^{2} + 2\sqrt{\pi} \, k z \, \mbox{\rm e}^{\textstyle -z^{2}} - 
\mbox{\rm e}^{\textstyle -2z^{2}}} \right) \, .
\end{equation}
In principle, this solves the problem: Eq.\ (7) gives $\sigma_{0}$ as 
$\sigma$ times a function of $N$ and $n$ only, and then Eq.\ (6) gives 
$\mu_{0}$ as a best estimate of the template. 

However, $\sigma$ is derived from a small number of points and has a
large statistical uncertainty. Through Eq.\ (6) these uncertainties
translate into noise in the template spectrum. Henrichs et
al. (\cite{HK94}) have shown that for IUE spectra an empirical
relation can be found between $\sigma_{0}$ and flux, which implies
that we can derive $\sigma_{0}$ from $\mu_{0}$ (see also Howarth \&
Smith \cite{HS95}). This relation is based on much better statistics
and gives more reliable results. Since $\mu_{0}$ is the quantity we
want to derive, we have to estimate $\sigma_{0}$ from $\mu$, so that
in principle the scheme has to be iterated. In practice, these
iterations are barely needed. The method was tested numerically by
selecting a given number of highest flux points from a gaussian
distribution (with known $\mu_{0}$ and $\sigma_{0}$) of randomly
chosen points. The method was found to be extremely powerful: for
$k=$ 0.01 we predicted $\mu_{0}$ within 2\%. The higher the value of
$k$, the better the obtained estimate for $\mu_{0}$ (and
$\sigma_{0}$), but we are limited by the fact that only a few points
per wavelength bin are not disturbed by intrinsic variability.

In Fig.\ 1 we present, as an example, the \siiv\ resonance doublet of
\cyg. From the available 149 spectra we selected the 5 highest flux
points per wavelength bin and computed the average $\mu$ and variance
$\sigma^{2}$. We show $\mu(\lambda)$ and $\sigma(\lambda)$ by a
dotted line in the upper and lower panel of Fig.\ 1, respectively. The
average flux points and standard deviations of all 149 spectra are
shown as a thin line. Note that the average of the 5 highest flux
points $\mu(\lambda)$ lies, as expected, systematically above the
template $\mu_{0}(\lambda)$, and that in the lower panel the scatter
in $\sigma$ is large. The derived $\mu_{0}(\lambda)$ and
$\sigma_{0}(\lambda)$ are represented by a thick line. We see that the
template $\mu_{0}(\lambda)$ and the average spectrum are identical in
the continuum regions, but that the template deviates significantly
from the average spectrum in the variable resonance line. In the lower
panel of Fig.\ 1 the similarity between the estimated $\sigma_{0}$
(thick line) and the measured variance (thin line) is obvious in the
wavelength regions outside the \siiv\ line.

Figs.\ 2 and 3 display the relevant parts of the derived template
spectra. Shown are the P~Cygni profiles including DACs. For
comparison, a representative observed spectrum (thin line) is shown in
the same panel.  The quality of the template can be checked by
searching for ``emission'' in the obtained quotient spectra (i.e.\
where the flux exceeds unity), which should be present if the template
spectrum still contains additional absorption with respect to some of
the observed spectra.  Another check is to compare the variance
$\sigma^{2}$ in the selected highest flux points in the variable line
with the value of $\sigma^{2}$ in a constant part of the spectrum with
a comparable exposure level. For example, in the case of \xip\ we were
forced on these grounds to select only the two highest flux
points. Although a logical consequence of our method is that a few
points (e.g.\ at least one per wavelength bin for \xip) in the
quotient spectra will exceed unity, we found that these points are not
randomly distributed among the quotient spectra, but are concentrated
in only a few of them.  This means that we can only improve on this by
enlarging the sample of observed spectra.  Fortunately, the
``emission'' in the quotient spectra of \xip\ is modest. For the other
stars this problem turns out to be even less important.

\begin{figure}[tbp]
\centerline{\psfig{figure=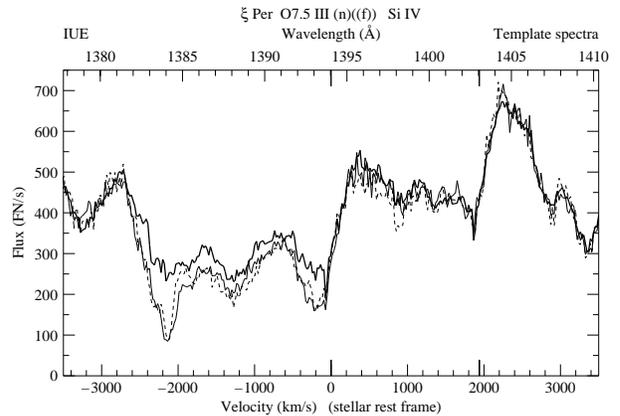,width=8cm}}
\caption[]{Comparison of templates in the \siiv\ region of $\xi$~Per
constructed from 1994 data (thin line), 1991 data (dashed line) and
all data prior to 1994 (thick line) used in the present analysis.}
\end{figure}

In order to investigate whether there exists such a thing as an
undisturbed wind profile which represents the state to which the wind
returns after the passage of a DAC, we constructed templates based on
the spectra of the individual datasets and compared them. The most
challenging case is that of \xip, where the template we used is based
on the two highest flux points out of a total of 124 spectra. As can
be expected, the resulting templates are identical (within the noise)
in the continuum regions of the spectrum, but differ in the absorption
troughs of the P~Cygni lines. Of the templates based on the individual
datasets prior to 1994, the October 1991 template (36 spectra) is
closest to the template we used, but clearly contains more absorption
in the blue doublet component (Fig. 4, \siiv\ doublet). We also
constructed a template from a dataset of 70 spectra of \xip\ collected
during a campaign lasting 10 days in October 1994 (Henrichs et
al. \cite{HD98}), which are not comprised in the sample studied in
this paper. The October 1994 template resembles the template we
constructed from all spectra until 1991 even more, but also here
additional absorption is found at high velocities. This demonstrates
that, given a sufficient timespan covered by the dataset, our method
is capable of finding the same underlying wind profile using different
datasets; in the case of \xip, however, only at low and intermediate
velocities. For the latter star we have to conclude that in October
1991 and October 1994 the high-velocity part of the wind has a larger
optical depth compared to other campaigns, which could be due to
long-term variability.

\subsection{Modeling DACs in quotient spectra}

The isolated DACs in a quotient spectrum, obtained after division of
an observed spectrum by the template, are modeled in the way described
in Henrichs et al. (\cite{HH83}, see also Telting \& Kaper
\cite{TK94}). The DACs are assumed to be formed by plane-parallel
slabs of material in the line of sight, giving rise to an absorption
component with central optical depth $\tau_{c}$ and (Doppler)
broadening parameter $v_{t}$. The intensity of the component is
described by:
\begin{equation}
I(v) = \mbox{\rm e}^{\textstyle - \tau_{c} \phi(v)}
\end{equation}
with the Gaussian profile function 
\begin{equation}
\phi(v) = \exp\bigg[ {-\left( \frac{v - v_{c}}{v_{t}} \right)^{2}} \bigg]
\end{equation}
where $v$ is the velocity with respect to the stellar rest frame and
$v_{c}$ the Doppler displacement at line center. With the (reasonable)
assumption that the displacement $v_{c}$ and broadening parameter
$v_{t}$ are identical for the absorption components in both lines of
the resonance doublet, the DACs in both doublet components can be
modeled simultaneously knowing the doublet separation
$v_{\mbox{\scriptsize split}}$ and the ratio in oscillator
strength. This leaves only three free parameters for each pair of
DACs. These parameters ($v_{c}$, $v_{t}$, and $\tau_{c}$) are
determined by means of a $\chi^{2}$-method in order to obtain the best
fit.  The signal-to-noise ratio in each point of the quotient spectrum
is estimated using the empirically determined dependence of this ratio
on the flux (Paper~I). The $\chi^{2}$ criterion then gives the formal
errors in the derived parameters (Telting \& Kaper \cite{TK94}). We
fitted our model to spectral data within the velocity range $-$3000 to
$+$2500 \kms\ with respect to the principal (short-wavelength) doublet
component.  From these parameters we determine the column density
$N_{\mbox{\scriptsize col}}$ associated with the absorption components
(Henrichs et al. \cite{HH83}):
\begin{equation}
N_{\rm col} = \frac{m_{{\mbox\rm e}}c}{\pi e^{2}}
\frac{\sqrt{\pi}}{f \lambda_{0}}
\frac{\tau_{c} v_{t}}{(1 + v_{c}/c)}
\end{equation}

\begin{figure}[tbp]
\centerline{\psfig{figure=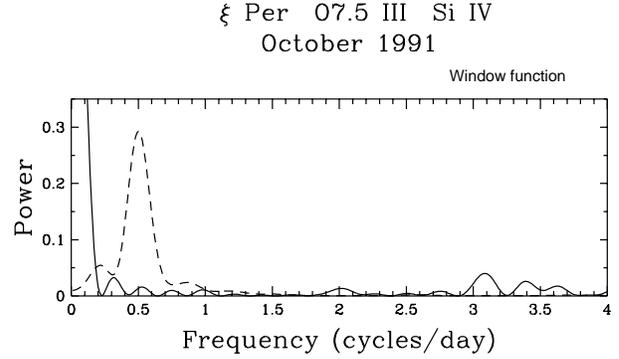,width=8cm}}
\caption[]{The window function obtained for the time series of \xip\
in October 1991 (full line) is very clean thanks to the continuous
coverage. For comparison, the cleaned power spectrum, integrated over
the red absorption component of the \siiv\ doublet, is shown as a
dashed line.}
\end{figure}

As an example, we show in Figs.\ 2 and 3 model fits (thick line) to
quotient spectra which were obtained after division of a
representative spectrum (thin line, left-hand panel) by the template
(thick line). In most cases, a number of components were modeled
simultaneously, with a maximum of 4 DACs in one doublet
component. Furthermore, all fit parameters were kept free. As an
initial condition we used the results of the previous fit and modeled
the spectra in a sequence determined by the time of
observation. Sometimes we had to go back and forth through the dataset
when it appeared that we had missed the first (weak) signs of the
development of a new DAC at low velocity. In some cases we had to
correct for bad normalization in the quotient spectrum by allowing for
a very weak and wide (sometimes a few thousand \kms) absorption
component. This way we were able to determine the central velocity
and width of the other components with reasonable accuracy. In these
cases the derived column densities are, however, less accurate.

\begin{figure*}[tbp]
\vspace{19.4cm}
\caption[]{\xip\ O7.5~III(n)((f)): An overview of the DAC behaviour in
the \siiv\ resonance lines over a period of 4 years. The quotient
spectra (middle panel) are shown as a ``dynamic'' spectrum in the
grey-scale panel below. Outstanding is the similarity of the DAC
patterns, though detailed changes do occur. The significance of
variability (Paper~I) is indicated in the upper panels (thick line),
together with the mean spectrum of that particular year (thin line).}
\end{figure*}
 
\begin{figure*}[tbp]
\centerline{\psfig{figure=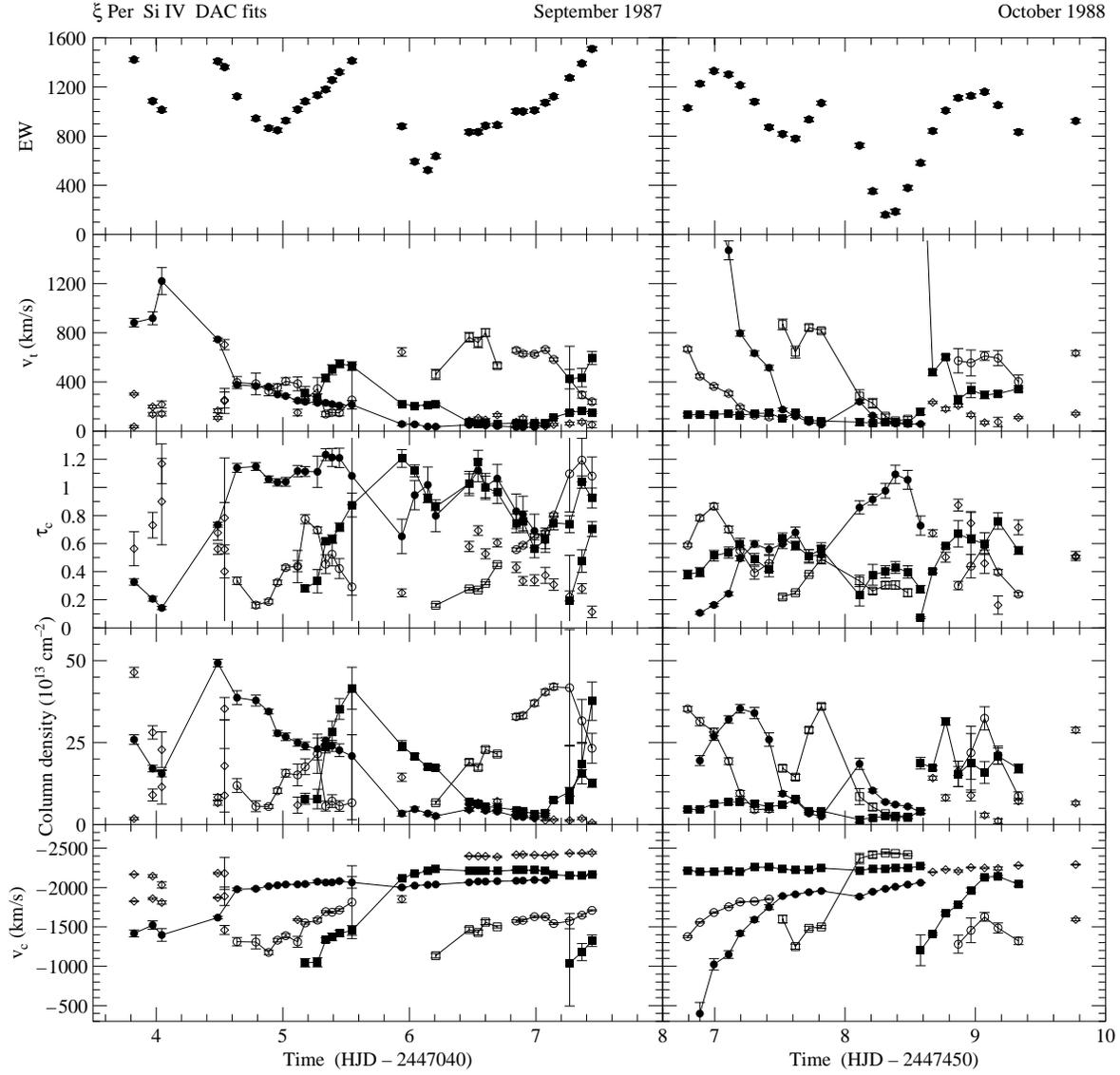,height=15cm}}
\caption[]{\xip\ September 1987 and October 1988: Parameters of DACs
in the \siiv\ resonance doublet. The central velocity $v_{c}$, central
optical depth $\tau_{c}$, and width $v_{t}$ of the components are
plotted as a function of time (in days). The figures belonging to
datasets of the same star are plotted on identical scale. The column
density $N_{\mbox{\scriptsize col}}$ is calculated with Eq.\ (10) and
expressed in units of 10$^{13}$ cm$^{-2}$. In the top panel the total
equivalent width (EW) of the \siiv\ quotient spectra is given in
\kms. The points suggested to belong to the same DAC event are
interconnected.}
\end{figure*}

\begin{figure*}[tbp]
\centerline{\psfig{figure=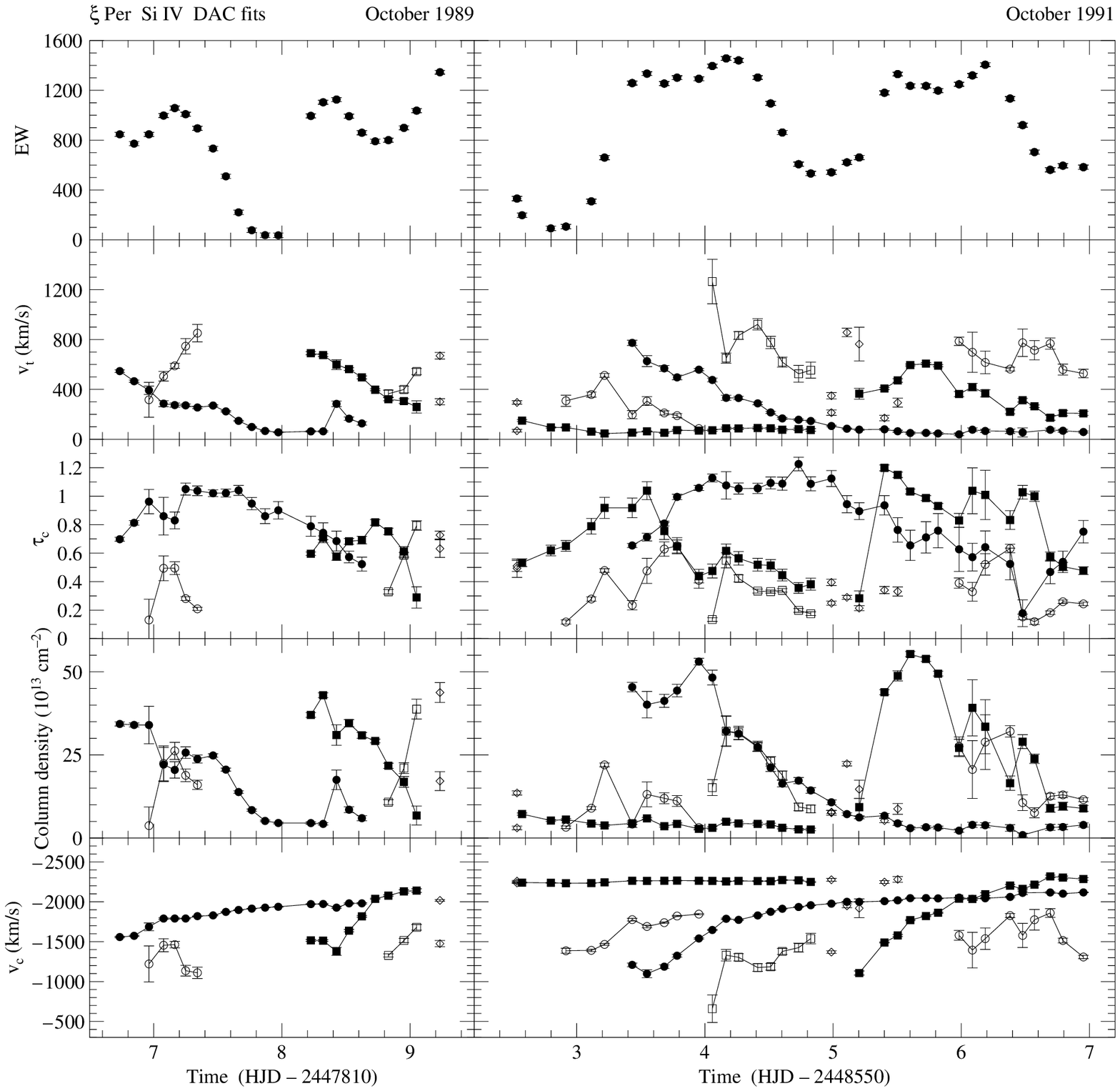,height=15cm}}
\caption[]{\xip\ October 1989 and October 1991 (as Fig.\ 7): DACs in the
\siiv\ resonance doublet.}
\end{figure*}

\subsection{Period-search analysis}

We also inspected the P~Cygni lines for the occurrence of {\it
periodic} variability. The spectral lines were normalized to the level
of the surrounding continuum through division by a first-order
polynomial. The period-finding algorithm we applied, consists of an
ordinary Fourier transformation for non-equidistant temporal sampling,
followed by a CLEAN stage in which the window function, which is due
to incomplete temporal sampling of the stellar signal, is iteratively
removed from the Fourier spectrum (Roberts et al.  \cite{RL87}). The
continuous observations from space result in a very nice window
function which does not include a peak at a period of one day (see
Fig.~5 for an example). The window function was removed in 400
iterations with a gain of 0.2. The studied frequency range runs from
0.01 to 10 cycle~day$^{-1}$, with a frequency step of 0.01 c/d. A
Fourier spectrum is computed for each wavelength bin (0.1~\AA). The
power in the 2-dimensional periodograms resulting from the CLEAN
analyses can, provided that the window function could be
satisfactorily removed, be transformed back to amplitudes in continuum
units (amplitude = 2$\, (\mbox{power})^{1/2}$), contrary to the summed
power in the summed 1-dimensional periodograms. More details on the
used method and references can be found in Gies \& Kullavanijaya
(\cite{GK88}) and Telting \& Schrijvers (\cite{TS97}).

It turns out that for several of our targets the derived period is
close to, or sometimes even larger than the time span $\triangle t$
covered by our observations. In these cases we can only conclude that
a long-term variation is present in the data, which may or may not be
periodic. We have indicated these ``periods'' by putting them between
square brackets (e.g. [6.2$\pm$1.8~d]).

An outcome of the period-search analysis is the phase information. For
a given period, the corresponding sinusoid's phase (in our figures in
units of radians) is known in each wavelength bin. For example,
periodic features moving from the red to the blue side of a spectral
line will give rise to a shift in phase with wavelength. Owocki et
al. (\cite{OC95}) used the observed phase diagram of a periodic
modulation in UV resonance lines of the B0.5~Ib star HD~64760 as
supporting evidence for the occurrence of corotating streams in the
wind. In this particular example, however, the so-called ``phase
bowing'' did not correspond to the evolution of the, though present,
DACs, but to (sinusoidal) modulations in flux at low and intermediate
velocities (Fullerton et al. \cite{FM97}). It would be very
interesting to know whether these findings also apply to the here
studied O-star wind variability.

\section{DAC behaviour in individual stars}

In this section we describe the results based on the modeling of DACs
in the UV P~Cygni lines and period-search analyses for each star
separately. For some stars several time series are available, so that
we can study the behaviour of DACs also on a longer (yearly)
timescale. We have chosen to analyse in detail only those time series
(from the sample presented in Paper~I) that exhibit significant
variability and cover a sufficiently long time interval.

In most cases the \siiv\ resonance doublet was used to model the DACs;
the other UV resonance lines (\nv\ and \civ) are often saturated. The
two main-sequence stars in our sample, \mon\ and \lac, are variable in
the \nv\ (and \civ) profiles, but not in the \siiv\ doublet, which is
too weak to show any wind features. For the supergiants \cep\
and \zor\ we modeled the DAC behaviour in both the \siiv\ and the
\nv\ doublet. The subordinate \niv\ line at 1718~\AA\ of \xip, \sao,
and \lab\ (for the latter also the \heii\ line at 1640~\AA) varies in
concert with the DACs in the \siiv\ doublet, but at low velocities
only. In the wind profiles of \cam\ we did not detect any significant
variations, except for some marginal variability in the blue edge of
the \siiv\ and \civ\ lines (Paper~I).

To facilitate a visual comparison, the figures showing the time
dependence of the DAC parameters are plotted on scale. The scale of
the time axis is identical in all figures. For a given star, the scale
of the vertical axes is the same for each dataset. 

\begin{figure}[tbp]
\vspace{10cm}
\caption[]{\xip\ October 1991: Two-dimensional periodogram of the
\siiv\ doublet at 1400~\AA. The CLEANed Fourier spectrum obtained in
each wavelength bin (horizontal axis) in the \siiv\ line is plotted as
a function of frequency (in cycles~day$^{-1}$, vertical axis). The
power is represented in levels of grey (white cut: 10$^{-5}$; black
cut: 10$^{-3}$). The mean \siiv\ profile is plotted in the bottom
panel. For example, at a wavelength of 1388~\AA\ maximum power is
detected at a frequency of 0.5 c/d. An important result is that the
Fourier spectrum is different at the highest velocities with a
dominant peak at lower frequency. Also note that the power shows a gap
at the asymptotic DAC velocity.}
\end{figure}

\begin{figure}[tbp]
\centerline{\psfig{figure=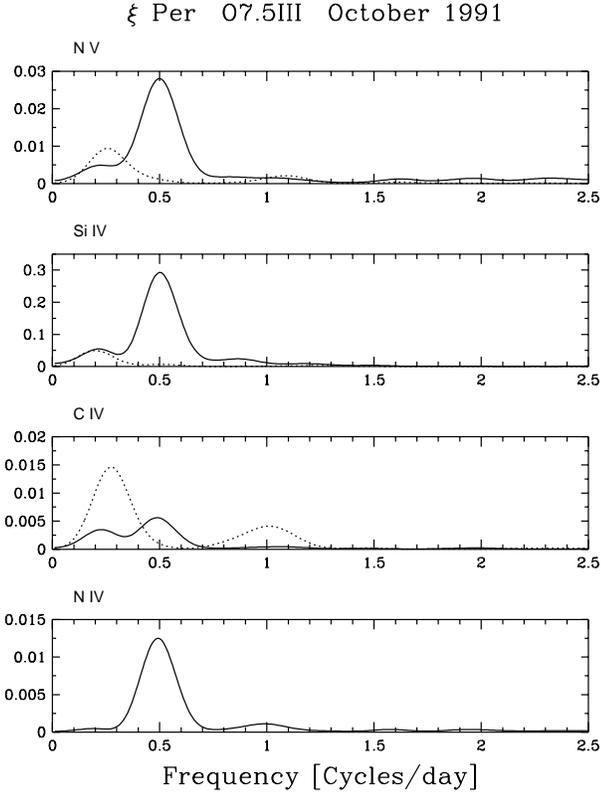,width=8cm}}
\caption[]{\xip\ October 1991: One-dimensional power spectra (obtained
by integrating a 2d-periodogram such as displayed in Fig.~9 over a
given wavelength range) of the \nv, \siiv, \civ, and \niv\ P~Cygni
lines. The periodic variations at low velocity are represented by a
solid line, the variations at the blue edge by a dotted line. The
2-day period (0.5 c/d) appears in all lines at low velocity, while the
blue edge seems to vary with approximately twice that period.}
\end{figure}  

\subsection{HD 24912 (\xip) O7.5 III(n)((f))}

We observed \xip\ four times in the period 1987--1991. Fig.~6 presents
an overview of the DAC behaviour in the \siiv\ resonance doublet and
enables a visual comparison to be made between the different
datasets. As discussed in Paper~I, the variability pattern is
characteristic for a given star, but shows detailed changes from year
to year. The timescale of variability (i.e.\ the recurrence timescale
of DACs) and the range in velocity over which variability is observed,
remain the same. The strength of the absorption components, however,
varies from event to event.

A minimum-absorption template was constructed from the 124 \siiv\
spectra of \xip, which was used to generate quotient spectra
(Fig.~6). The isolated DACs were fitted to a model profile as
described in section 2.2; in Figs.\ 7 and 8 the evolution of the DAC
parameters ($v_{c}$, $v_{t}$ and $\tau_{c}$) are presented for the
different sets of spectra. In the upper panels of Figs.\ 7 and 8 the
total equivalent width (integrated over velocity rather than
wavelength) of the quotient spectra is given as a function of
time. The error bars represent 1$\sigma$-errors (cf.\ Telting \& Kaper
\cite{TK94}). We connected the points that we identify as belonging to
the same DAC event. The criteria for selection of these points are
based on the assumption of continuity in the three fit parameters and
the column density.

\begin{figure}[!ht]
\centerline{\psfig{figure=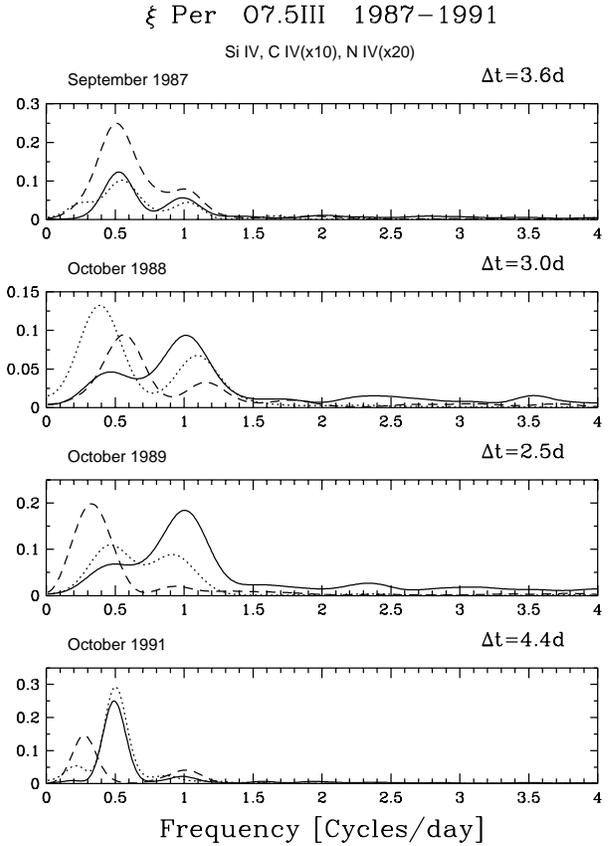,width=8cm}}
\caption[]{\xip\ 1987--1991: A comparison of the (summed) 1d power
spectra calculated for the different datasets. The solid line
represents the integrated power ($\times$20) for the \niv\ line
($-$1000 to 0 \kms), the dotted line shows the power in the red
doublet component of the \siiv\ profile ($-$1800 to 0 \kms), and the
dashed line gives the power ($\times$10) in the blue edge of the
\civ\ doublet ($-$2950 to $-$2650 \kms).}
\end{figure}

\begin{figure}[tbp]
\vspace{10cm}
\caption[]{\xip\ 1987 and 1991: Phase (in radians) of the 2.0-day
period as a function of position in the line. The phase is only
plotted when the summed power is larger than 0.0001. The 1987 points
are displayed as crosses, the 1991 points as filled circles. The two
phase diagrams are matched at an arbitrarily chosen velocity for
comparison. Phase bowing is clearly observed in the \siiv\ doublet.}
\end{figure}

The derived column density of a DAC first increases and then
decreases; a maximum is reached at a velocity of about $-$1500
\kms. The maximum in $N_{\mbox{\scriptsize col}}$ is 
5$\times$10$^{14}$ cm$^{-2}$.  The other DAC parameters vary 
less strictly. The
maximum width of a strong component is about 750 \kms\ just after its
appearance at low ($\sim -$1000 \kms) velocity. We have modeled narrow
components at their terminal velocity having widths smaller than 50
\kms, i.e.\ nearly down to the spectral resolution of the IUE
spectrograph, but not as narrow as the DACs observed with the GHRS
onboard HST (Shore et al.\ \cite{SA93}).  The maximum central optical
depth of a DAC that is observed for \xip\ is
$\tau_{c}^{\mbox{\scriptsize max}} \sim$1.2.

Two different types of DACs can be discriminated in the \siiv\ doublet
of \xip: (1) strong absorption components that remain visible for more
than two days (filled symbols) and (2) more short-lived, sometimes
suddenly disappearing DACs (open symbols), preceding or following a
strong component within half a day.

For each dataset we performed a period analysis on the \nv, \siiv, and
\civ\ resonance lines and the \niv\ subordinate line. The Fourier
analyses clearly detect periodicity, the cleanest periodogram being
obtained for the October 1991 dataset (see Fig.~9 for the \siiv\
results), when \xip\ was covered for the longest period of time
($\triangle t =$ 4.4~d; consequently, all periods derived from this
dataset that are close to or longer than 4.4~day are not well
determined and will be printed between square brackets). In each
wavelength bin a CLEANed Fourier spectrum is calculated and plotted as
a function of frequency (in cycles~day$^{-1}$), the power being
represented in levels of grey. Periodic variability is detected in the
blue-shifted absorption troughs of the \siiv\ doublet (also in the
other studied lines). The dominant period is found at 2.0$\pm$0.2~day,
 almost over the full width of the blue-shifted absorption
profile. As a conservative estimate of the accuracy of the period we
take the half-width of the peak in the (summed) power spectrum. The
2-day period is clearly reflected by the change in total EW with time
(Figs. 7 and 8).

At higher velocities ($>$1800 \kms), however, a peak is found at a
longer period: in the \siiv\ line at [5.0$\pm$1.5]~day, which is
longer than the length of the observing run and therefore
unreliable. In the other lines (\nv\ and \civ) also a longer period is
detected at the blue edge (between $-$2950 and $-$2350 \kms) of the
profiles, at [3.8$\pm$1.0] and [3.6$\pm$0.9]~day, respectively
(Fig.~10, dotted lines). These (saturated!) resonance lines also
include the 2-day period at lower velocities (ranging from about
$-$2000 to 0 \kms with respect to the rest wavelength of the red
doublet component). The longer period is in all three cases consistent
with a period of 4 days, i.e.\ twice the period observed at lower
velocities. The subordinate \niv\ line only exhibits the 2.0$\pm$0.3~day 
period, at velocities between $-$1000 and 0 \kms.

The 2-day period is also present in spectra from previous
campaigns. In Fig.~11 we compare 1d-power spectra from different lines
for the period 1987--1991. The solid line shows the power spectrum of
the \niv\ line, integrated over the absorption part of the line from
$-$1000 to 0 \kms. The campaigns covering the shortest period of time
(1988 and 1989) result in a higher weight to a 1-day period (1.0$\pm$0.1~d).
 The evidence for a longer period in the blue edge of the
profiles, such as found in 1991, is not very strong. This is not
surprising given the short time span covered by these observations.



\begin{figure*}[!ht]
\begin{center}
\centerline{\psfig{figure=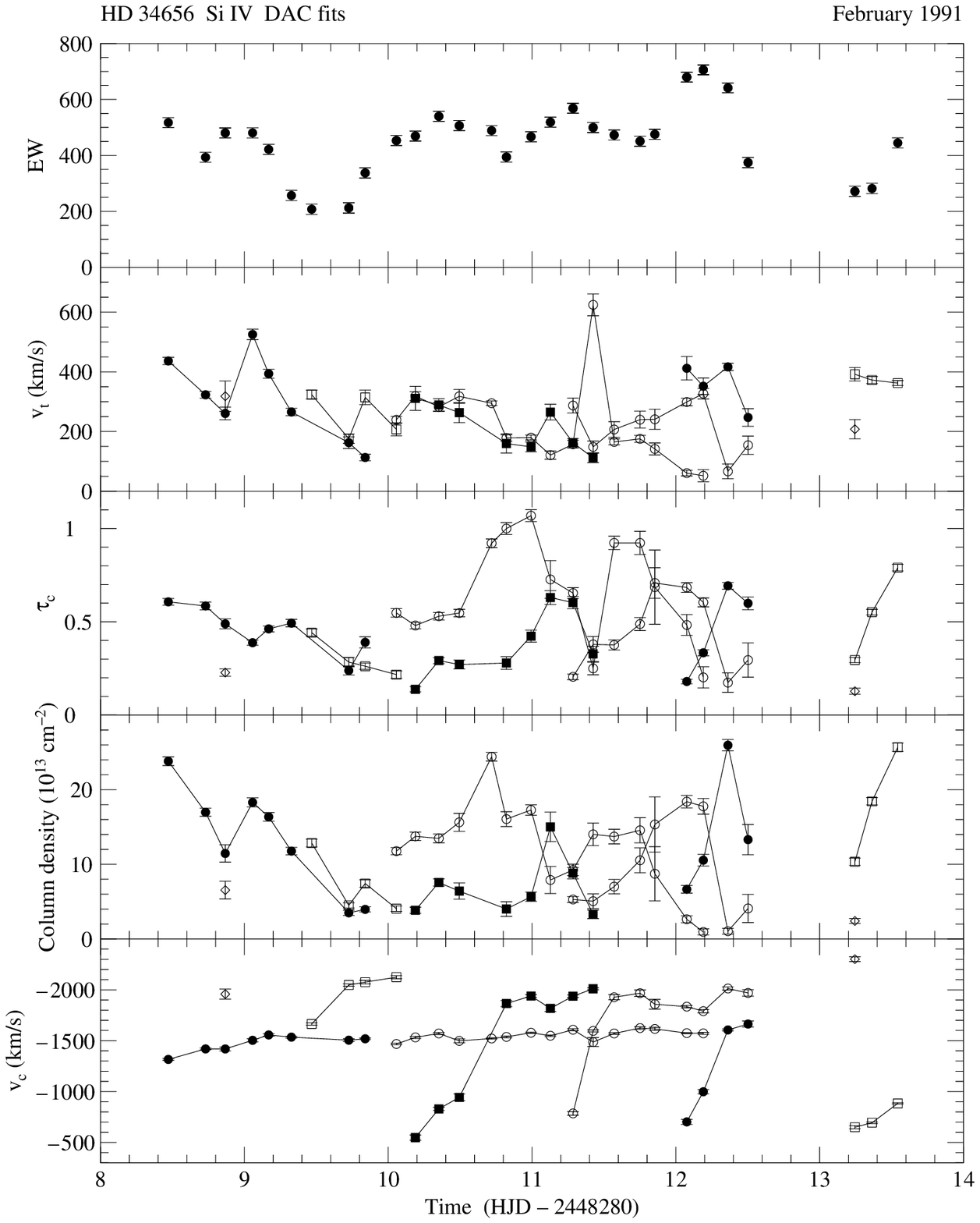,height=15cm}}
\caption[]{As Fig.\ 7: parameters of DACs in the \siiv\ doublet at 
1400~\AA\ of \sao\ O7~II(f) in February 1991. The DAC behaviour can be
interpreted in terms of a ``slow'' and a ``fast'' pattern, the latter
with a recurrence timescale close to one day.}
\end{center}
\end{figure*}

Why do the observed variations at higher velocities exhibit a longer
period, about twice as long as at lower velocities? If our
interpretation is correct, this factor of two difference in
periodicity is due to the peculiar DAC behaviour in the wind lines of
\xip. In Paper~I we pointed out that in this star strong DACs reach
different asymptotic velocities, leading to crossings of successive
components. We found that the DAC's asymptotic velocity alternates
between 2050 \kms\ and 2275 \kms\ (in Figs.~7 and 8 the high-velocity
DAC is indicated by filled squares). Therefore, only one out of two
DACs would reach the far blue edge of the profile resulting in a
frequency of once every four days at the highest velocities.

It turns out that if one wants to understand the edge variability, a
detailed knowledge of the DAC behaviour is essential. The recognition
that the asymptotic velocity of a DAC can have two different values
has important consequences for the use of this diagnostic to determine
the terminal velocity of a stellar wind (Henrichs et al. \cite{HK88},
Prinja et al. \cite{PB90}). We assume here that the highest value
corresponds to \vinf.  Furthermore, the alternating behaviour of DACs
in the wind of \xip\ suggests that a full cycle includes two strong
DAC events, i.e.\ 4 days.

Fig.~12 shows the phase of the 2.0-day period as computed by the
Fourier analysis. Only the results are plotted for the 1987 (crosses)
and 1991 (filled circles) datasets; the 1988 and 1989 datasets give
similar results, although there a 1-day period is dominant (the phase
diagrams are, however, almost identical to those of the 2-day
period). The phase (in radians) is plotted as a function of position
in the line (measured in velocity with respect to the blue doublet
component) for the three resonance lines and the \niv\ subordinate
line. The phase has only a physical meaning if at the given velocity
the period is detected. We plot the phase for points with power
exceeding 10$^{-4}$, which is just above the level of the continuum in
the power spectrum (Fig.~10). The error in phase does not follow
straightforwardly from the Fourier analysis; a possibility is to bin
the data and use the spread in phase as an indication of the error. We
did not include error bars in the phase diagrams, but the spread in
the points indicates that the error on the phase must be small.

The \siiv\ lines clearly demonstrate the so-called phase bowing (cf.\
Owocki et al. \cite{OC95}, Fullerton et al. \cite{FM97}), indicative
of curved wind structures (like corotating interaction regions) moving
through the line of sight. A maximum in phase is reached at about
$-$1200~\kms\ in the \siiv\ profile. In the CIR model, this velocity
corresponds to the velocity of the material in the spiral-shape CIR
first moving out of the line of sight. The maximum phase difference
$\triangle \phi$ (measured between $\approx -$1000 and $-$2000 \kms)
is about one radian, which is comparable to that observed for the B
supergiant HD 64760 ($\triangle \phi \approx$ 0.6 $\pi$~radians,
Fullerton et al. \cite{FM97}). Note that a decrease in phase towards
more (less) negative velocities corresponds to an accelerating
(decelerating) feature moving upwards in time (as in the grey-scale
plot of e.g.\ Fig.~6). Both datasets show a similar picture; for
comparison, an arbitrary constant was added. The \niv\ line shows the
rising phase only at low velocities, consistent with the behaviour of
the \siiv\ (and \nv) doublet. Saturation of the \civ\ and \nv\ lines
causes the absence of points bluewards of $-$1000 \kms. At low
velocities the 1987 points seem to indicate a faster increase in
phase. Another small difference between 1987 and 1991 is seen at high
velocities and in the blue edge. In 1987 the edge variability is very
pronounced and leads to a signal at the 2-day period around $-$2500
\kms\ in the resonance lines as mentioned earlier.

\subsection{HD 34656 O7 II(f)}

\begin{figure}[tbp]
\centerline{\psfig{figure=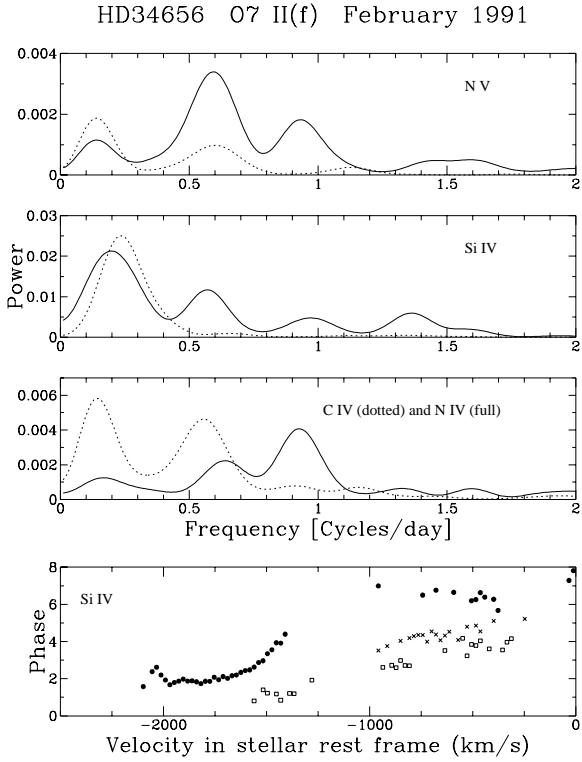,width=8cm}}
\caption[]{HD 34656 February 1991: the top three panels present
1d-periodograms illustrating the periodic variability in the \nv,
\siiv, \civ, and \niv\ P~Cygni lines. The solid lines show the power
summed over the low-velocity ($<$1500 \kms) part of the profiles, the
dotted lines represent the high-velocity part. A [4.5-day] period is
only found at high velocities, a 1- and/or 2-day period dominates at
low velocity. The bottom panel shows the phase (in radians) of the [4.5]
(filled circles), 1.8 (open squares), and 1.1-day period (crosses) as
measured from the \siiv\ line (power $>$10$^{-4}$), only for the blue
doublet component ($\lambda_0 =$ 1393.755~\AA).}
\end{figure}

\begin{figure}[tbp]
\centerline{\psfig{figure=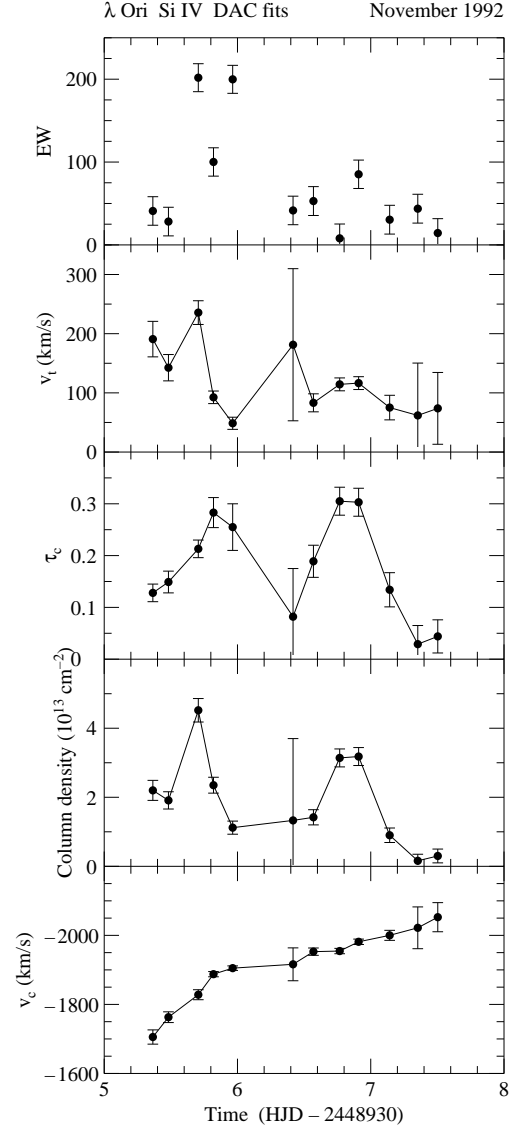,height=15cm}}
\caption[]{As Fig.\ 7: the parameters of the migrating DAC in the 
\siiv\ profile of the O8~III((f)) star \lor\ in November 1992.}
\end{figure}

\begin{figure*}[tbp]
\centerline{\psfig{figure=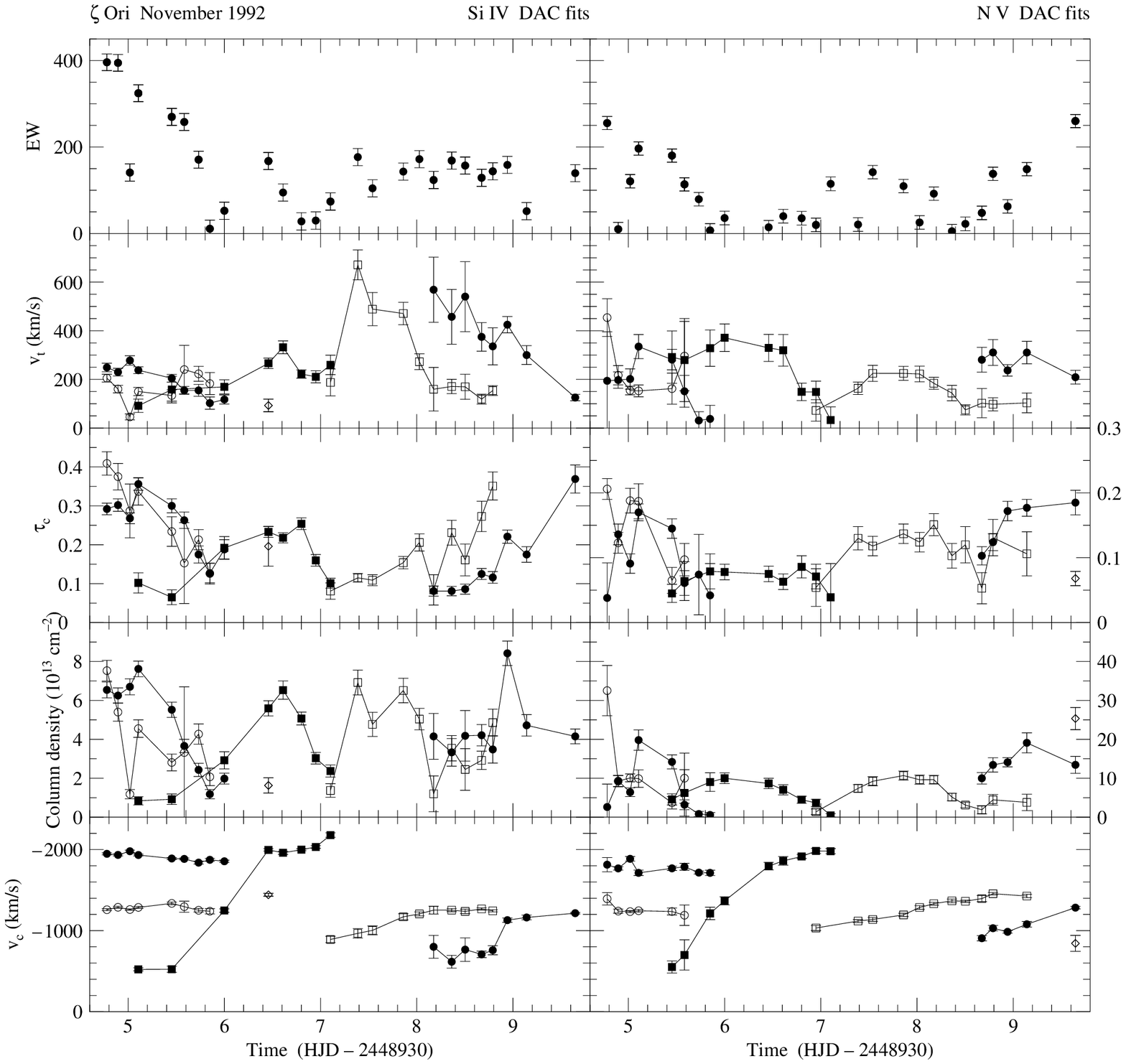,height=15cm}}
\caption[]{As Fig.\ 7: \zor\ O9.7~Ib November 1992: fit parameters of
DACs in the \nv\ (left panel) and \siiv\ profile (right panel). A new
DAC appears about every 1.5 days.}
\end{figure*}

We observed the O7 supergiant HD 34656 only once; the 27 spectra
obtained during 5.1 days in February 1991 show the rapid recurrence of
several DACs appearing at low velocity (about $-$500 \kms). In Fig.\
13 the fit parameters are shown for the DACs in the \siiv\ line. The
DAC behaviour in this star is quite complicated: DACs appear about
once every day (observable as regular humps in EW), but around day 10
the strength of the absorption at $-$1500 \kms\ increases rapidly. It
appears as if an additional (variable) absorption component is present
at $-$1500 \kms\ at all times. This makes it difficult to assign
points to specific DAC events.

The column density of this additional component suggests that it
belongs to a separate DAC event, reminiscent of a {\it persistent}
component as observed in the wind lines of \lor\ and \mon\ (see
below). At day 9.8 the $N_{\mbox{\scriptsize col}}$ of this component
reaches a minimum of 4$\times$10$^{13}$ cm$^{-2}$ and then starts to
strengthen rapidly to a maximum of 2.5$\times$10$^{14}$ cm$^{-2}$. We
suspect that at day 9.8 a new ``DAC'' replaces the former one at
$-$1500 \kms\ and remains at that velocity until day 12. This would
mean that the absorption component at $-$1500 \kms\ follows a
different cycle than the ``rapid'' DACs appearing once every day.  The
variations might be interpreted as caused by two DAC patterns
superimposed on each other: a ``slow'' pattern consisting of DACs with
an asymptotic velocity of $-$1500 \kms, and a ``fast'' pattern of DACs
that reach higher velocities ($\sim-$2150 \kms).  The rapid DAC
pattern is also (weakly) present in the \niv\ line, the DACs can be
traced down to a velocity of $-$200 \kms. The origin of the slow
pattern is not clear; it might reflect a variation in wind structure
on a larger spatial scale than the structures causing the DACs
belonging to the rapid pattern.

\begin{figure*}[!ht]
\centerline{\psfig{figure=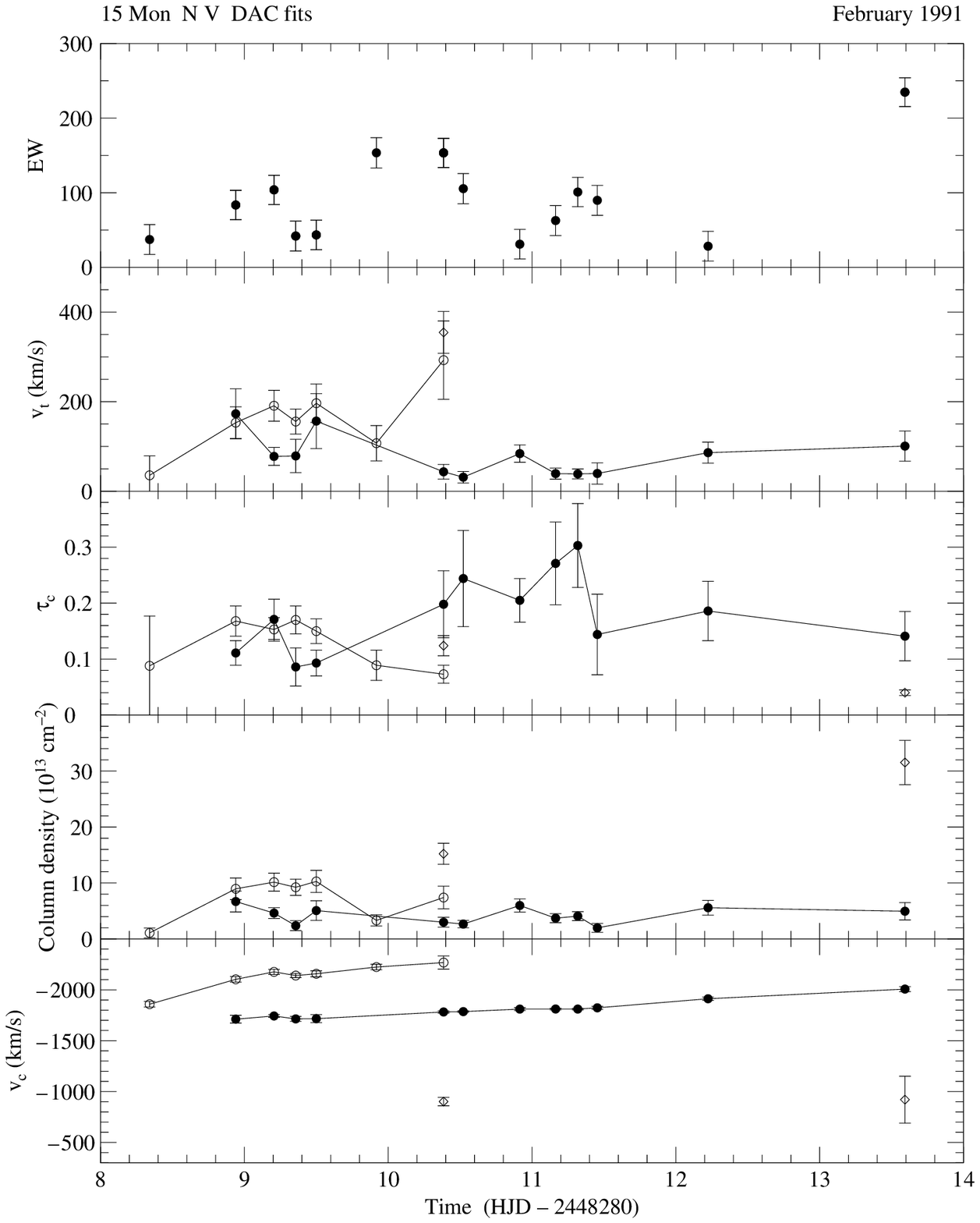,height=15cm}}
\caption[]{As Fig.\ 7: \mon\ O7 V((f)) February 1991: two DACs are
detected in the \nv\ time series. The DAC appearing at day 9 has a
width $v_{t} <$ 150 \kms; this narrow width has also been observed for
the DACs in \lor\ and \lac.}
\end{figure*}

Period analyses of the P~Cygni lines of HD 34656 also indicate a
bimodality in the DAC behaviour. Fig.~14 presents 1d-periodograms
obtained after summing the 2d-periodograms over a given wavelength
interval. The solid lines show the power detected at low velocity
($<$1500 \kms), the dotted lines the periodic behaviour at the blue
edge. The dominant period found in the \niv\ line is 1.1$\pm$0.1~d
and corresponds to the DAC recurrence timescale in the ``rapid''
pattern. This period is present in the low-velocity part of all four
lines. A period of 1.8$\pm$0.3~d is detected in all lines (at low
velocity), and might be related (alias) to the 1-day period
variation. At high velocity (dotted lines) a longer period dominates:
[4.5$\pm$1.8~d] (in the case of \siiv). This period might relate to the
slow pattern (see also the corresponding changes in EW, Fig. 13).

The phase (in radians) of the 3 detected periods is plotted as a
function of position in the \siiv\ line in the bottom panel of
Fig.~14. The phase is plotted for the blue doublet component only. In
this line, most power is measured for the 4.5-day period at velocities
bluewards of $-$1500~\kms. For all periods, the phase shows a
declining trend towards higher velocities, but no obvious phase
bowing, such as detected for \xip, seems to be present.
\subsection{HD 36861 (\lor) O8 III((f))}

We observed the O8~III((f)) star \lor\ during the November 1992
campaign and obtained 27 spectra ($\triangle t =$ 5.0~d).  The \siiv\
resonance lines of \lor\ consist of a sharp photospheric absorption
component, slightly asymmetric to the blue due to wind contamination,
plus an absorption component at $-$2000 \kms\ (Paper~I). Also the \nv\
and \civ\ resonance doublets exhibit this {\it persistent} absorption
feature at $-$2000 \kms, in both doublet components. We modeled the
quotient \siiv\ spectra of \lor\ (Fig.\ 15) and found one DAC
accelerating from $-$1700 to $-$2000 \kms, i.e.\ reaching the position
of the persistent component. The occurrence of the DAC is reflected by
the rise in \siiv\ EW around day 5.7, shown in the upper panel of
Fig.\ 15. We covered one DAC event, resulting in a lower limit of
about 5 days to the DAC recurrence timescale. These observations
suggest that the persistent component is regularly ``filled up'' by
new absorption components when they reach the terminal velocity of the
wind (which can, if this interpretation is valid, be determined from
the position of the persistent component).

\begin{figure*}[~ht]
\centerline{\psfig{figure=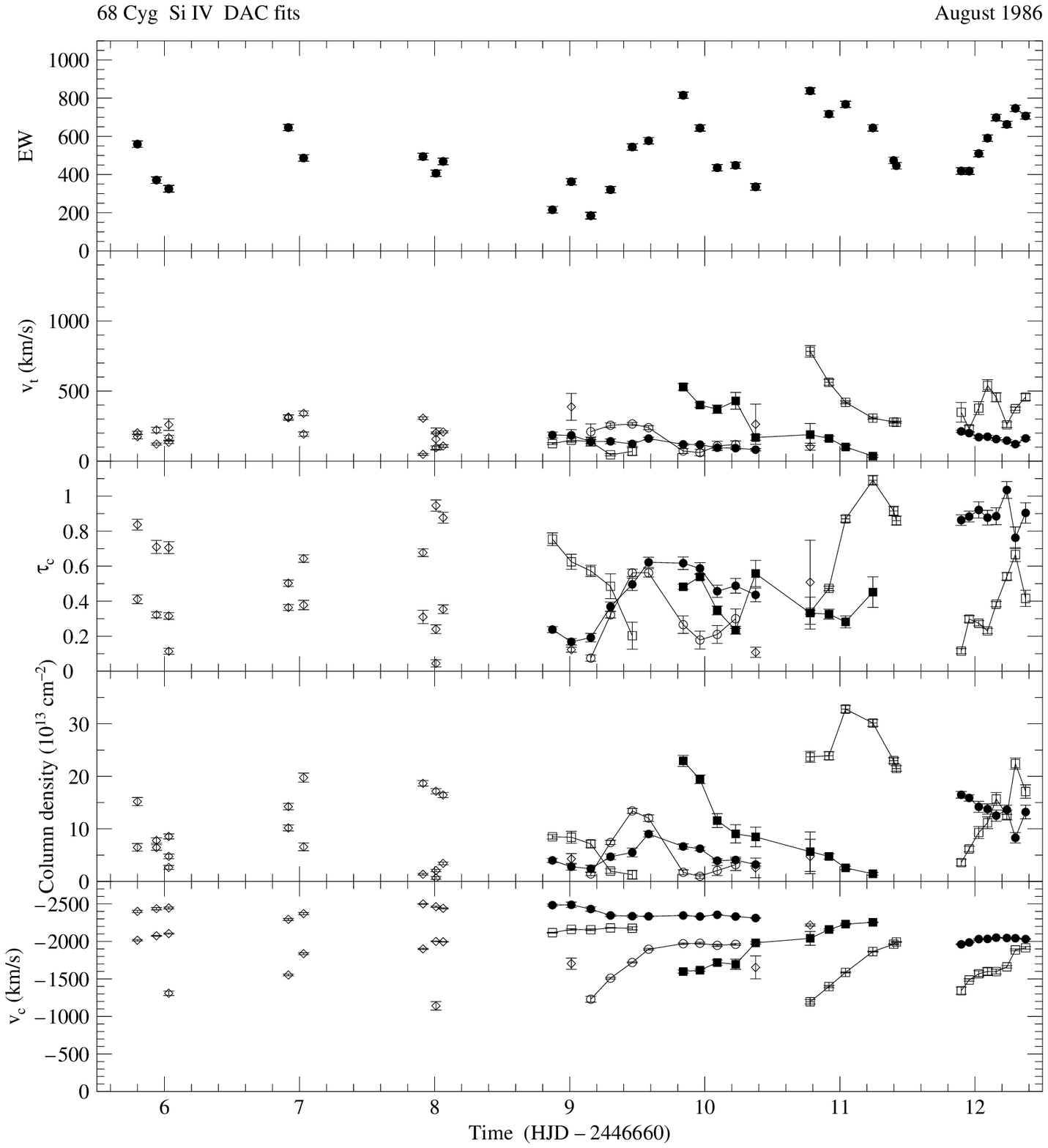,height=15cm}}
\caption[]{\cyg\ O7.5~III:n((f)) August 1986 (as Fig.\ 7): parameters
of DACs in the \siiv\ doublet. This dataset was also analyzed by
Prinja \& Howarth (\cite{PH88}). During the first three days time
coverage is very sparse, so that the different DACs could not be
labeled. Close inspection of the available datasets (cf.\ Figs. 19 and
20) results in the recognition of typical DAC patterns. On this basis
the DACs are labeled with symbols to highlight the pattern.}
\end{figure*}

Compared to DACs in other stars in our sample, the measured column
density is low, about 5$\times$10$^{13}$ cm$^{-2}$ in \siiv. The
central optical depth does not exceed 0.3 while the width of the
component is smaller than 250 \kms. Both in the \nv\ and in the \siiv\
line a period around 4 days is detected (between 1500 and 2000 \kms):
[3.4$\pm$0.8~d] and [4.7$\pm$1.4~d], respectively. Since
only one DAC event was observed, the physical significance of these
periods is doubtful. It might just reflect the presence (and
subsequent absence) of a DAC. 

\subsection{HD 37742 (\zor) O9.7 Ib}

\begin{figure*}[!ht]
\centerline{\psfig{figure=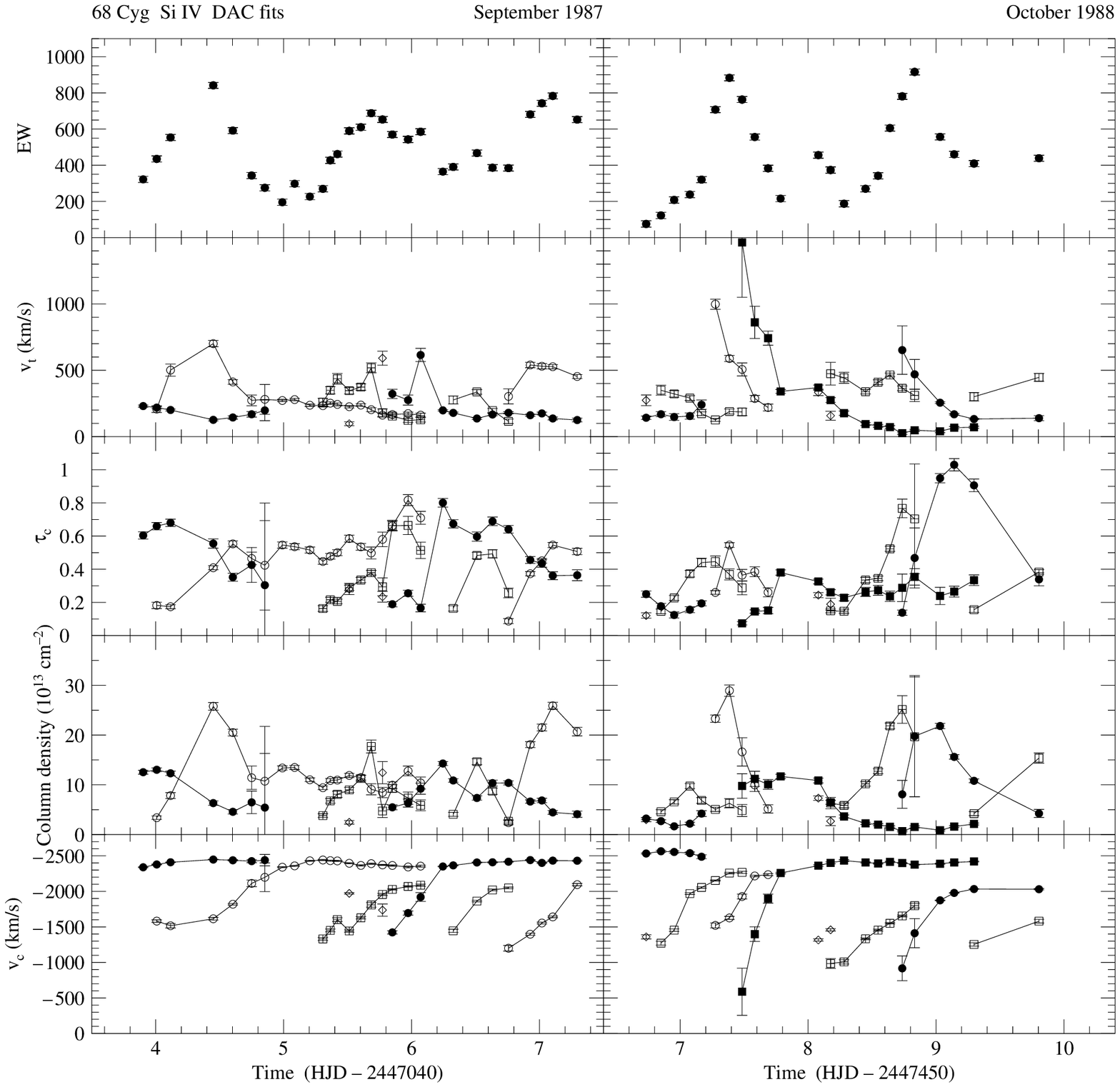,height=15cm}}
\caption[]{\cyg\ September 1987 and October 1988 (as Fig.\ 7): DAC
parameters derived from the \siiv\ doublet.}
\end{figure*}

Moving DACs are found in both the \nv\ and the \siiv\ resonance
doublet of the O9.7~Ib star \zor. The results from our fit procedure
are shown in Fig.\ 16. For this star it is very difficult to identify
the different DAC events. Our reconstruction is as follows: At day 5.4
and 7.1 the appearance of a DAC is registered in both lines, with the
difference that the first component accelerates faster than the second
one, and reaches a velocity of $-$2000 \kms; the second component ends
at a much lower velocity ($-$1250 \kms), similar to the component
present since the beginning of the observations. At day 8.7 a new
component develops, consistent with a recurrence timescale of about
1.6 days. (Alternatively, bearing in mind the $v_{c}$ versus time plot
(Fig. 16), the points at low ($-$500 \kms) velocity at day 5 could be
connected to the event starting at day 7 (open squares); the filled
squares would then be a continuation of the open and filled circles,
respectively.)

The Fourier analysis produces a peak in power at a 1.6$\pm$0.2~d
period; most of this power is concentrated in the velocity range
$-$1500 to $-$2150 \kms. The highest peak in the power spectrum is
found at a period of [$\sim$6] days, which is longer than the 5.1~d
span of the time series. H$\alpha$ observations of this star (Kaper et
al. \cite{KF98}) indicate a 6-day period as well, so that this period
might be real. The 1.6-day period is about a quarter of the 6-day
period.

The properties of the DACs in the two different lines can be
compared. The central velocities of the modeled components are similar
in both lines. The other DAC parameters show the same trend in both
the \nv\ and the \siiv\ line.  The column densities differ by a factor
of about two: maximum $N_{\mbox{\scriptsize col}} \sim$1.1$\times$10$^{14}$ 
and 0.6$\times$10$^{14}$ cm$^{-2}$ for \nv\ and \siiv,
respectively. The similarity of the DAC behaviour in the \nv\ and
\siiv\ lines supports the common view that the variable DACs reflect
changes in wind density (and/or the velocity gradient), rather than in
the ionization structure of the stellar wind.

\subsection{HD 47839 (\mon) O7 V((f))}

Two migrating DACs are found in the \nv\ doublet of the O7~V((f)) star
\mon. The first DAC is visible from the start of the campaign, the
second one appears just before day 9 (Fig.\ 17) and is very narrow
($v_{t} <$150 \kms). Such a narrow width is also observed for DACs in
P~Cygni lines of \lor\ and \lac. Because we have only 20 spectra in
total, it is difficult to construct a good template; this explains why
the EW of fits to some quotient spectra is negative (not shown).
Furthermore, the variations in the \nv\ profile have a small
amplitude, which makes the quality of the template an even more
important factor. The first DAC accelerates up to $-$2250 \kms, which
is considerably higher than the central velocity ($-$1900 \kms) of the
persistent component in the \nv\ (and \civ) profile. The second
component, which appears at a velocity of $-$1700 \kms, seems to join
the persistent component. The column density of the DACs is less than
10$^{14}$ cm$^{-2}$, but this is uncertain due to the poor
normalization.

On the basis of this dataset, only a lower limit of 4.5 days can be
set to the DAC recurrence timescale. The Fourier analysis indicates a
[6-day] period (power integrated from $-$1500 to $-$2350 \kms),
but the length of this period exceeds the 5.3~d observing period. 

\begin{figure*}[tbp]
\centerline{\psfig{figure=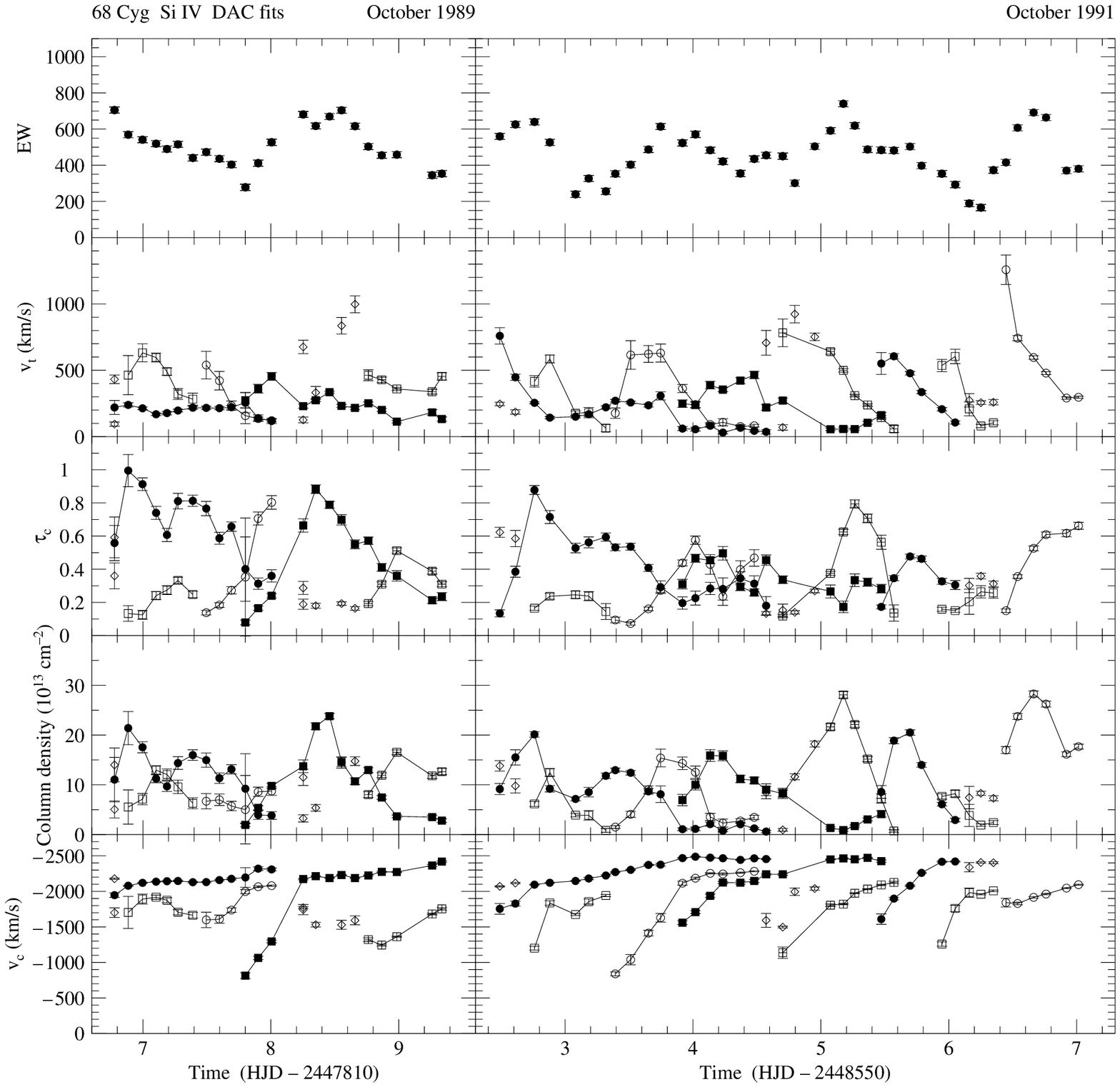,height=15cm}}
\caption[]{\cyg\ October 1989 and October 1991 (as Fig.\ 7): DAC
parameters in the \siiv\ doublet. The October 1991 dataset has the
longest time coverage (4.5~d) and includes 40 spectra.}
\end{figure*}

\subsection{HD 203064 (\cyg) O7.5 III:n((f))}

The August 1986 dataset of the O7.5~III:n((f)) star \cyg\ has been
analyzed by Prinja \& Howarth (\cite{PH88}). In Fig.\ 18 we show the
results of profile fits to this set of 33 spectra. Our results are
compatible with those of Prinja \& Howarth. In Fig.\ 19 the DAC
parameters derived for the September 1987 and October 1988 campaigns
are presented. The central velocities of DACs in the September 1987
spectra were presented by Fullerton et al. (\cite{FB91}). The DAC fits
of the October 1989 and October 1991 datasets are shown in Fig.\ 20,
the latter dataset having the longest time coverage with good time
resolution (5 days); we describe these observations first.

\begin{figure}[!ht]
\centerline{\psfig{figure=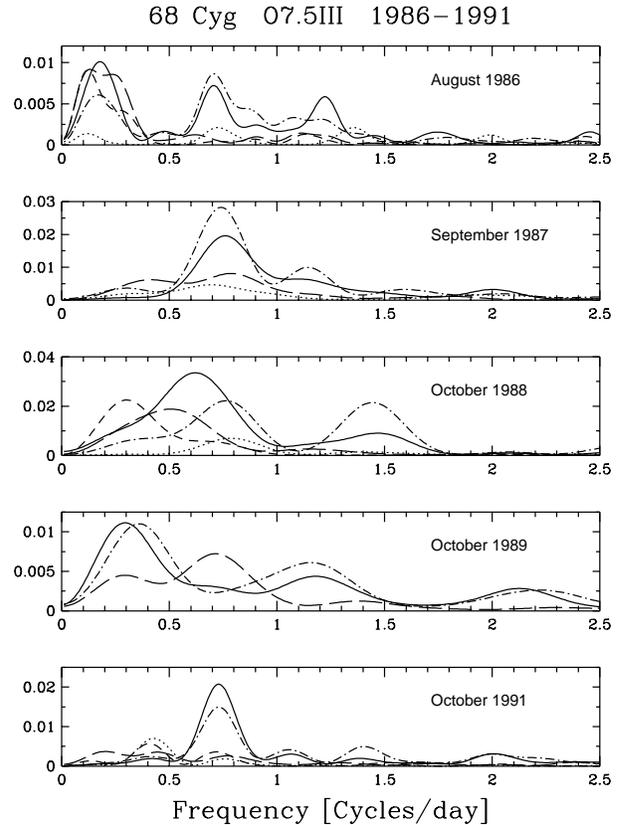,width=8cm}}
\caption[]{\cyg\ 1986--1991: A comparison of the (summed) 1d power
spectra calculated for the different datasets. The solid line
represents the summed power for the blue absorption component of
the \siiv\ line ($-$2100 to $-$600 \kms), the dashed-dotted line that
for the red doublet component ($-$2100 to $-$150 \kms). As expected,
both lines follow the same trend. The long-dashed line demonstrates
the periodicity found at the high-velocity edge of the \siiv\ profile
($-$2750 to $-$2100 \kms). The dotted line gives the integrated power
spectrum obtained for the red doublet component of the \nv\ profile
($-$1875 to $-$675 \kms). The short-dashed line shows the power spectrum
in the blue edge of the \civ\ doublet ($-$2800 to $-$2550 \kms).}
\end{figure}

\begin{figure}[!ht]
\centerline{\psfig{figure=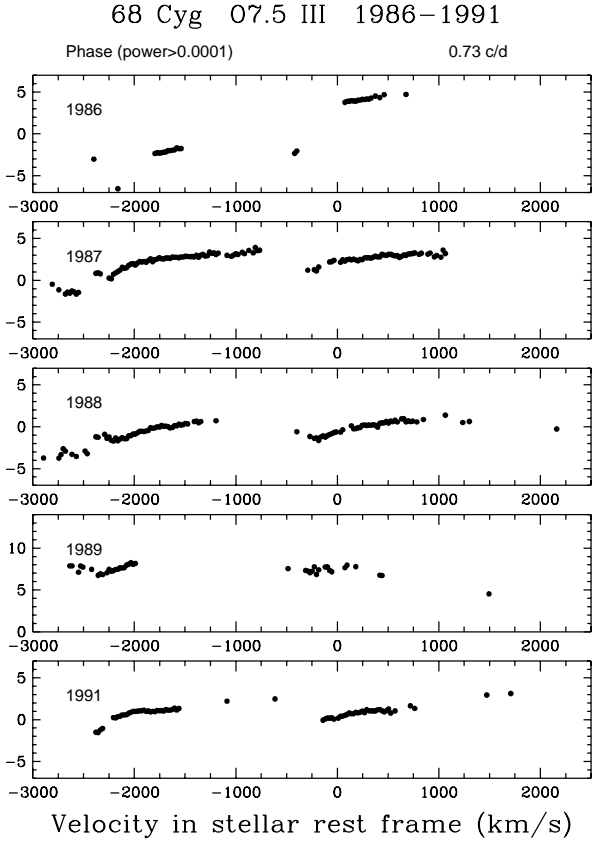,width=8cm}}
\caption[]{As Fig. 12: Phase diagrams for the 1.4-day period (0.73
c/d) detected in the \siiv\ lines of \cyg\ (phase in radians). The
datasets from different years give very similar results. Phase bowing
is not convincingly detected, this might be due to the absence of
variability at low velocities.}
\end{figure}

The Fourier analysis results in a pronounced peak in the power
spectrum at a frequency of 0.73$\pm$0.19~d$^{-1}$, corresponding to
a period of 1.4$\pm$0.2 day (Fig.~21). Assuming that this period is
the DAC recurrence time scale, we can identify {\it two} different
series of DACs moving through the \siiv\ profile (Fig. 20): series A,
which appears at JD 52.8, 53.9, 55.5, 56.5, and series B, appearing at
JD 53.4, 54.7, 56.0 (the sampling time of the UV data is 0.1 day). The
time lag between the two series is about half a day, series A followed
by B. Furthermore, it turns out that DACs belonging to different
series, reach different asymptotic velocities: series A accelerates to
a velocity of about $-$2450~\kms, which is $\sim$250~\kms\ higher
than the velocity reached by series B. Thus, like in the case of \xip,
the DAC's asymptotic velocity shows an alternating behaviour. An
important difference is, however, that in \xip\ the time span between
two successive DACs with different asymptotic velocity (i.e.\
belonging to different series) is equal to the DAC recurrence
timescale, while in \cyg\ the time lag between series A and B is about
0.5 days, i.e. not equal to the DAC recurrence timescale (1.4 days).

Simultaneous \ha\ observations indicate that incipient \ha\ emission
is observed prior to the appearance of DACs belonging to series A
(Kaper et al. \cite{KH97}), and not in between series A and
B. Alternatively, each DAC event in \cyg\ might consist of two
separate components, the event starting with an absorption component
from series A, followed half a day later by one from series B.

When comparing the different datasets, typical DAC patterns are
recognized (see also the figures in Paper~I). In Paper~I we 
pointed out the remarkable similarity between the DAC pattern in the
\siiv\ time series obtained in 1986 and 1988. In the figures we
labeled the DACs showing similar behaviour with the same symbol. Like
for \xip, the DAC behaviour suggests that the full cycle of
variability is twice the DAC recurrence timescale, which is 2.8
days. In exceptional cases (as, for example, in the time series of
September 1987 demonstrated by the absence of filled squares) a DAC
seems to be missing.  This is another indication that the general DAC
behaviour is very regular over the years, but that small though
significant differences do occur. We note that with our fit procedure
it is difficult to resolve two overlapping absorption components of
comparable strength, and that such double features will sometimes fit
as one single component, which causes unavoidable discontinuities in
the time sequence of the fit parameters.

The total DAC EW (upper panels) varies roughly with the 1.4~day
period.  The column density of the DACs reaches a maximum when the
component is half-way its acceleration (i.e.\ at a central velocity
around $-$1750 \kms), and then decreases with time.  The strongest
DACs reach a column density of 3$\times$10$^{14}$ cm$^{-2}$. Some
components remain visible for two days. The maximum central
optical depth $\tau_{c}$ is about 1, and the width of some DACs
exceeds 1000 \kms\ when they first appear at low velocity.

\begin{figure*}[!ht]
\centerline{\psfig{figure=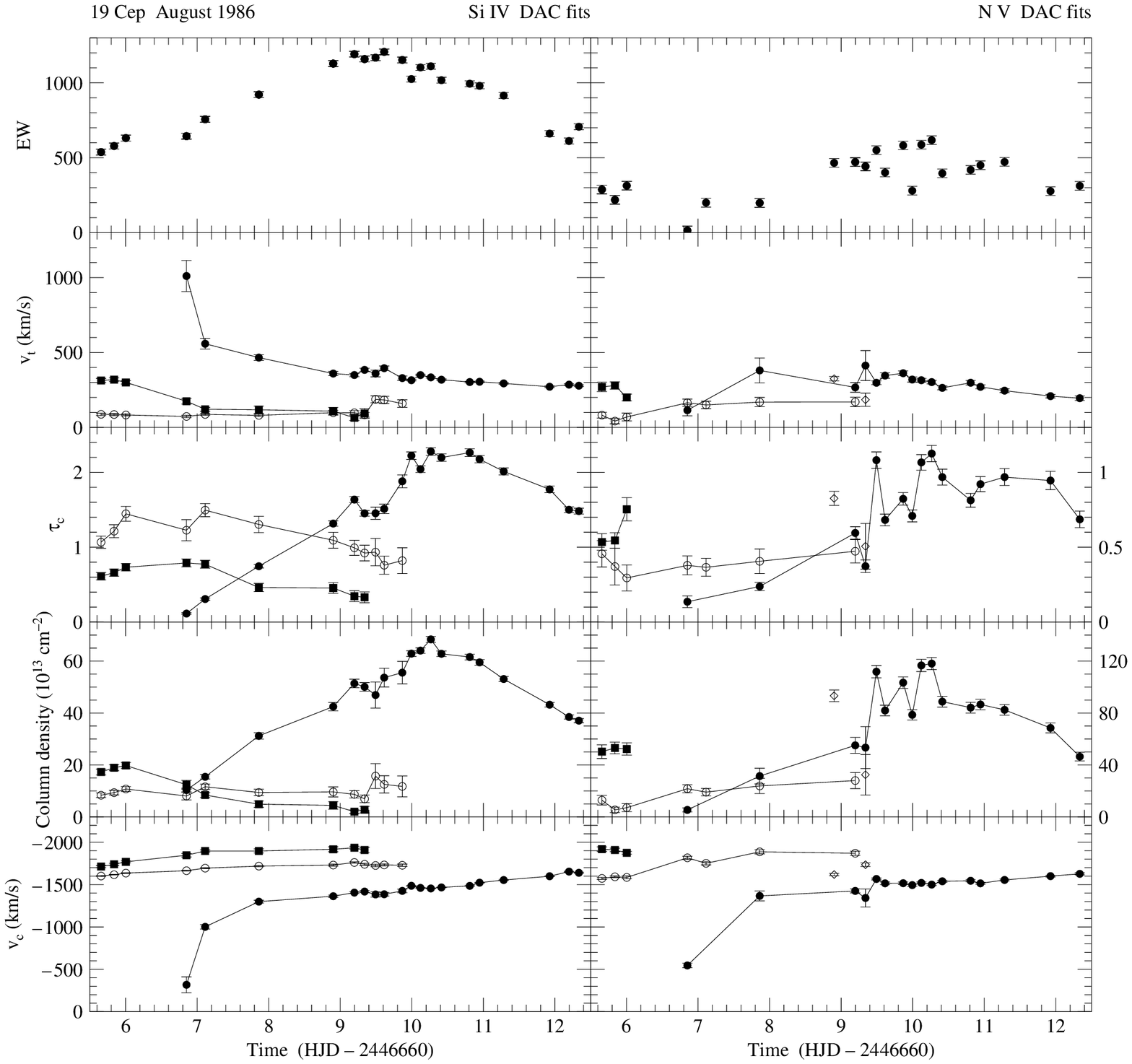,height=15cm}}
\caption[]{\cep\ O9.5~Ib August 1986 (as Fig.\ 7): DAC model
parameters for the \siiv\ (left) and nv\ (right) doublets. The two
profiles give consistent results and show slowly evolving DACs. The
DAC appearing at day 7 is the strongest component we encountered in
our O~star dataset.}
\end{figure*}

Fig.\ 21 presents the (summed) 1d power spectra obtained for different
UV P~Cygni lines in the period 1986--1991. The power diagrams are quite
complicated: at a given velocity several periods are detected, which
gradually change from one velocity bin to the other. Therefore,
excluding the October 1991 dataset, the summed 1d power spectra
contain several relatively broad peaks. It is not possible to
transform the summed power back to a variability amplitude in
continuum units. In all years, a period of 1.4 days is present, the
campaigns covering the shortest period of time (1988, 3.1~d, and 1989,
2.6~d) provide less significant results. The October 1991 dataset
includes a 2.4$\pm$0.4~d period in the high-velocity edges of the
profiles, but it remains uncertain whether this period could be
identified with the suggested 2.8~d full cycle of wind variability. The
other datasets produce power at periods around 3 days, but the
significance of any peak in power at a period close to 2.8~days is not
high.

The phase diagrams for the 1.4-day period (phase in radians) are very
similar from year to year (Fig.~22). The differences are mainly due to
changes in the amplitude of the variability (the phases are only plotted for
velocity bins with power $>$10$^{-4}$). The September 1987 dataset shows
the phase behaviour over the largest velocity domain. Phase bowing is
not convincingly detected, but this might be due to the absence of
variability at lower velocities (also the \niv\ line is not variable,
contrary to the \niv\ line in \xip).

\subsection{HD 209975 (\cep) O9.5 Ib}

\begin{figure*}[tbp]
\centerline{\psfig{figure=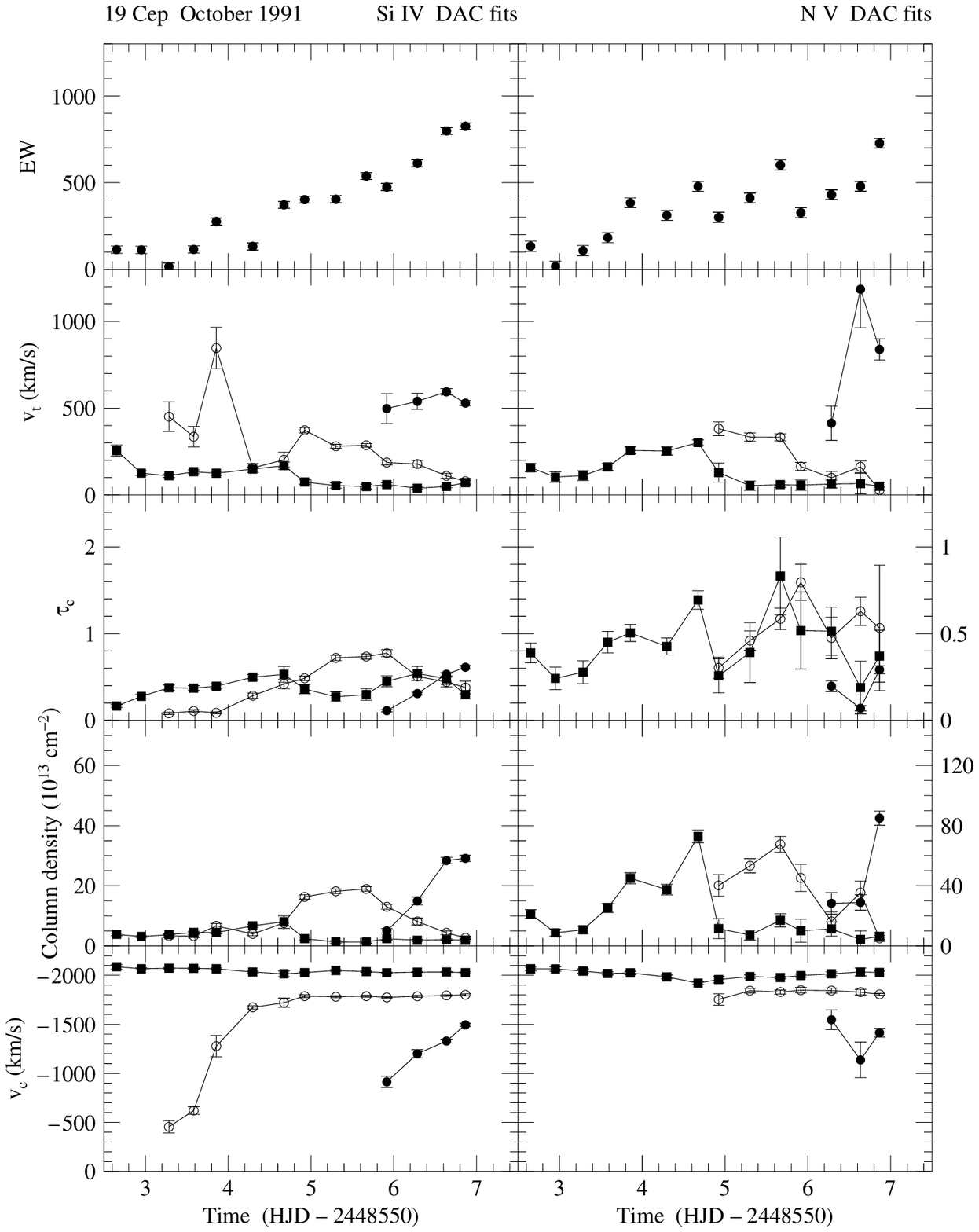,height=15cm}}
\caption[]{\cep\ October 1991 (as Fig.\ 7): DAC model parameters for
the \siiv\ (left) and \nv\ (right) profiles. Two DACs develop within a
time span of 2.7 days.}
\end{figure*}

\begin{figure*}[tbp]
\centerline{\psfig{figure=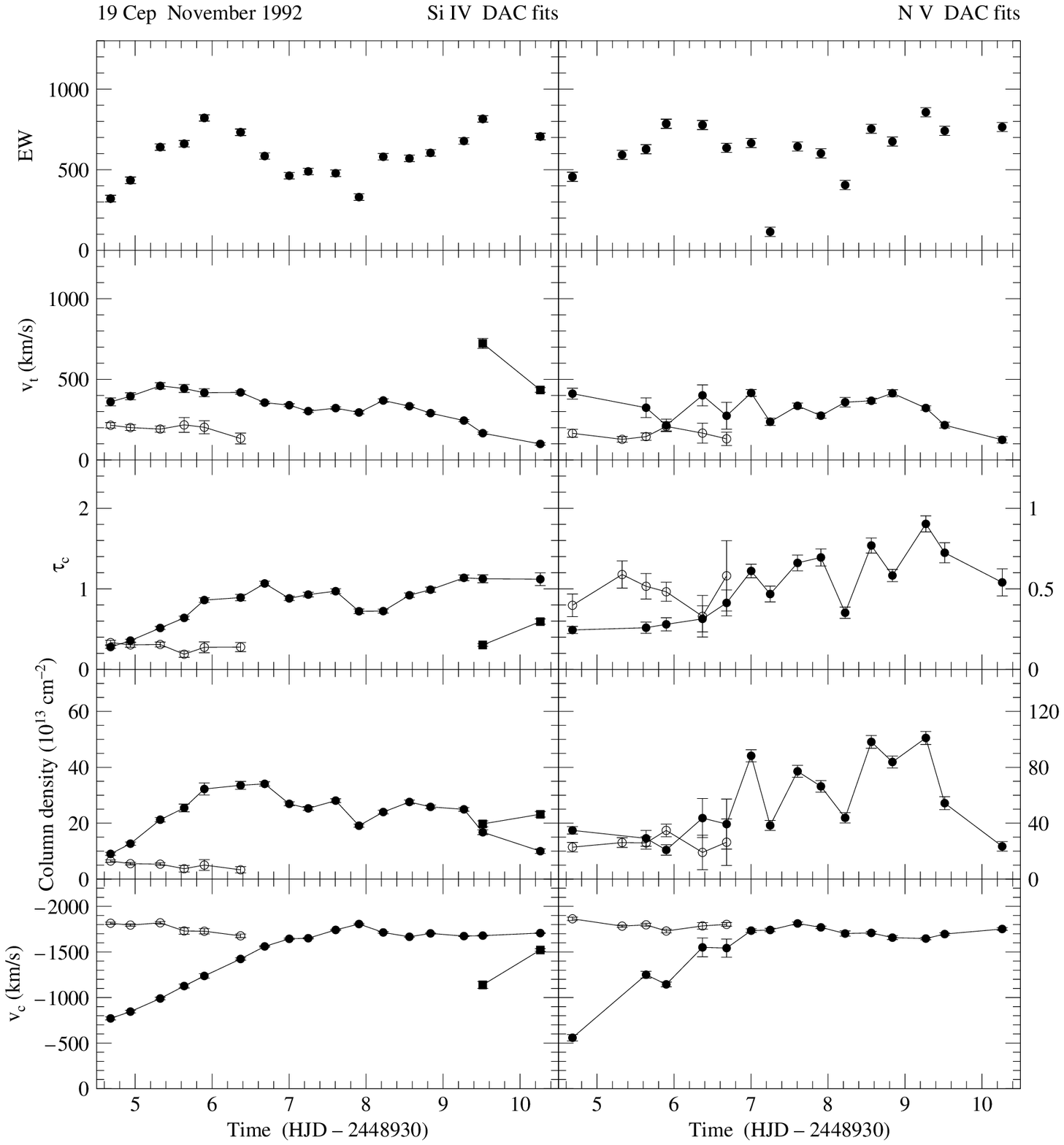,height=15cm}}
\caption[]{\cep\ November 1992 (as Fig.\ 7): DAC model parameters for
the \siiv\ (left) and \nv\ (right) profiles. The time interval between
the successive development of DACs is about 4.5 days.}
\end{figure*}

The O9.5~Ib star \cep\ provides a very clear picture of the DAC
phenomenon. Because the characteristic timescale of variability is
much longer (about 5 days) than in some other well-studied stars, the
behaviour of DACs is easily recognized. In Fig.\ 23 we present the
time evolution of the different fit parameters for the \siiv\ and
\nv\ spectra obtained in August 1986. The model fits to quotient
spectra of both resonance lines give comparable results. In August
1986 we detected three DAC events. A very strong absorption component
appears at low velocity ($v_{c}=-$300 \kms\ and $v_{t}=$ 1000 \kms) at
day 6.8. The column density of this component reaches a maximum of
6.8$\times$10$^{14}$ cm$^{-2}$ at a central velocity of $-$1450 \kms\
(in \siiv). This is the strongest DAC we encountered in our collection
of O-star spectra. The maximum central optical depth exceeds 2 at the
peak column density. The asymptotic velocity of the two weak
components present during the first half of the campaign is $-$1900
and $-$1750 \kms\ for the first and second component, respectively.

The time series obtained in September 1987 ($\Delta t =$ 3.2~d) and
October 1988 ($\Delta t =$ 3.8~d) spanned only a few days and did not
reveal the clear evolution of a DAC (Paper~I), which can be expected
on statistical grounds when the DAC recurrence timescale is on the
order of five days. It could also indicate that the strength of the
DACs varies from year to year.  In October 1991 (Fig.\ 24) a weak
component is present at its final velocity of $-$2050 \kms; at day 3.2
a new DAC appears at low velocity. The maximum column density reached
by this component is 1.9$\times$10$^{14}$ cm$^{-2}$ and
$\tau_{c}^{\mbox{\scriptsize max}}=$ 0.7. At day 5.9 a new DAC occurs with
rapidly growing strength. The time interval of 2.7 days is short if we
realize that in August 1986 the DAC recurrence timescale would be
estimated to be on the order of 5 days. The November 1992 observations
(Fig.\ 25) again show the development of a DAC, a fairly strong one
with maximum $N_{\mbox{\scriptsize col}}$ of 3.5$\times$10$^{14}$
cm$^{-2}$. At the end of this run (day 9.5) a new DAC appears in the
\siiv\ line, from which we again would conclude that the recurrence
timescale is about five days. The 2.7 days observed in October 1991
might correspond (within the errors) to half this period. 

The Fourier analysis of the October 1991 campaign does not
produce a significant peak in power at a period close to 2.5 days. All
three studied datasets include a long period, always close to or even
longer than the length of the campaign: [7.0$\pm$2.3~d] (1986,
$\Delta t=$ 6.7~d), [6.3$\pm$2.1~d] (1991, $\Delta t=$ 4.2~d), and 
[4.5$\pm$1.9~d] (1992, $\Delta t=$ 5.6~d) in the \siiv\ profile. 
The latter period
should get the highest weight. Fourier analysis of the \nv\ line
produces similar results. The obtained periods are consistent with the
period of 5 days suggested above. Also the H$\alpha$ observations that
were carried out simultaneously with the October 1991 IUE campaign
indicate a 5-day period (Kaper et al. \cite{KH97}).


\subsection{HD 210839 (\lab) O6 I(n)fp}

\begin{figure*}[tbp]
\centerline{\psfig{figure=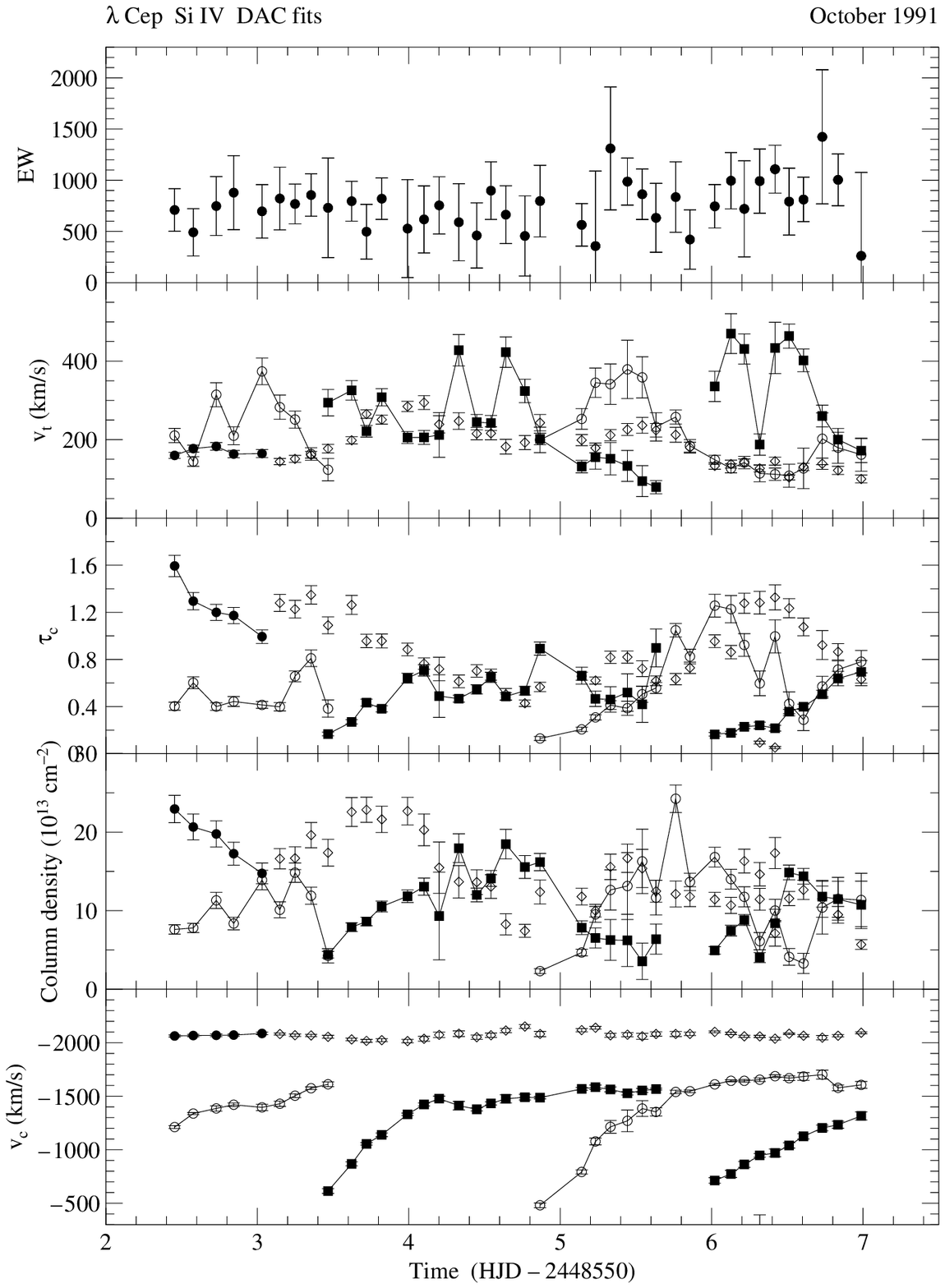,height=15cm}}
\caption[]{\lab\ O6~I(n)fp in October 1991 (as Fig.\ 7): DAC
parameters resulting from profile fitting of the \siiv\ doublet. The
velocity domain from $-$1600 to $-$2000 \kms\ is disturbed by large
fluctuations, because of saturation of the absorption profile.}
\end{figure*}

\begin{figure*}[!ht]
\vspace{10cm}
\caption[]{\lab\ O6~I(n)fp in October 1991: Grey-scale representations
showing the time evolution of the \siiv\ resonance doublet and the
\heii\ and \niv\ subordinate lines. A template spectrum (upper
panels) was used to produce residual spectra showing the development
and acceleration of 4 DAC events with better contrast. The variability
amplitude is illustrated by a thick line in the upper panel. The first
signs of DACs are observed at a velocity of $-$200	 \kms\ in these
subordinate lines.}
\end{figure*}

\begin{figure}[!ht]
\centerline{\psfig{figure=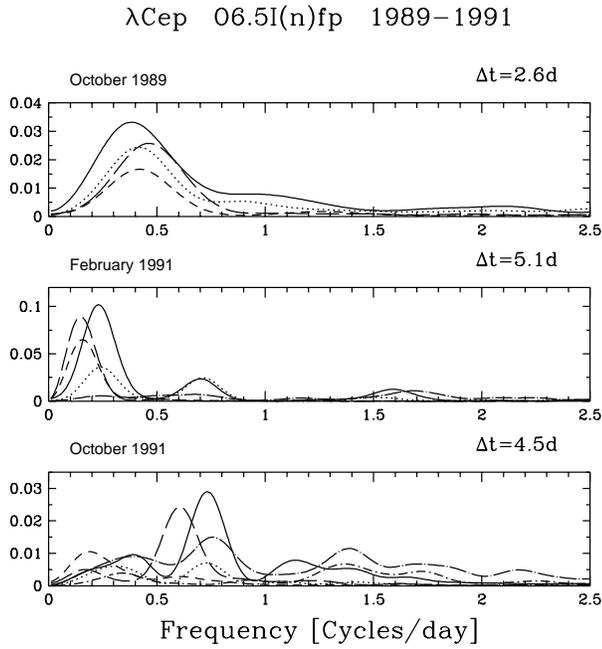,width=8cm}}
\caption[]{\lab\ O6~I(n)fp 1989--1991: Summed 1d power spectra
showing the periodical variability in the wind lines. The periodicity
in the \siiv\ line is measured in three intervals: the red and blue
doublet component (solid line: $-$600 to $-$1600 and dotted line:
$-$200 to $-$1400 \kms, respectively) and the high-velocity edge
(long-dashed line: $-$2200 to $-$2650 \kms). The \civ\ edge is
represented by the short-dashed line ($-$2200 to $-$2750 \kms). The
periodograms obtained for the subordinate \niv\ (long-dashed-dotted
line: $-$450 to $-$1500 \kms) and \heii\ (short-dashed-dotted line:
$-$200 to $-$350 \kms) lines are shown as well.}
\end{figure}

\begin{figure}[!ht]
\centerline{\psfig{figure=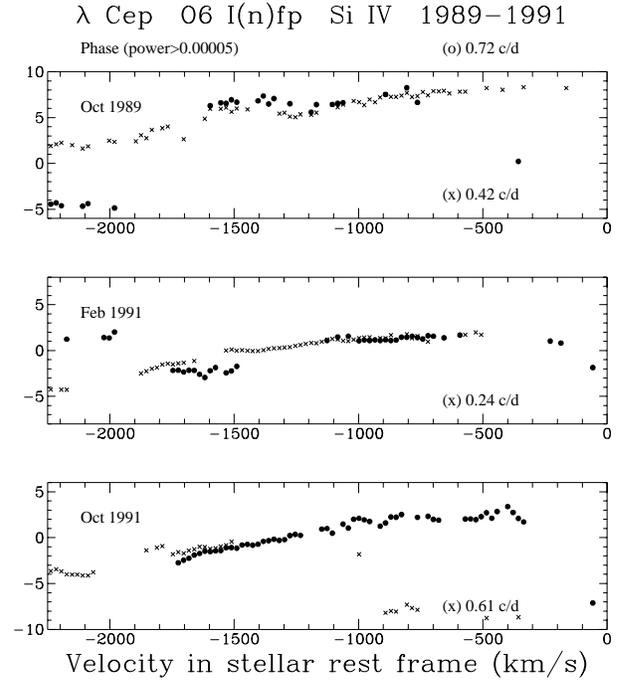,width=8cm}}
\caption[]{Phase diagrams obtained for the 1.4-day period (filled
circles) in the \siiv\ doublet of \lab; only the red doublet component
is shown ($\lambda_0 =$ 1402.77~\AA).}
\end{figure}

Although the O6~I(n)fp star \lab\ was monitored during six campaigns,
we present here the DAC modeling results only for the last campaign in
October 1991. The saturation of the UV resonance lines makes the
modeling of DACs very difficult. We performed a Fourier analysis on
the October 1989, February 1991, and October 1991 datasets. For the
October 1991 observations of the \siiv\ doublet we succeeded in
measuring the central velocity of DACs and were able to determine the
recurrence timescale. The velocity domain from $-$1600 to $-$2000
\kms\ is, however, difficult to model, since the quotient spectra
fluctuate with large amplitude due to the division by small numbers
and the low signal-to-noise. The DACs in the \siiv\ doublet (Fig.~26)
behave similarly as observed in other O~stars: the column density
increases after the appearance of a DAC and reaches a maximum (in this
case 2.5$\times$10$^{14}$ cm$^{-2}$) when a central velocity of about
$-$1500 \kms\ is reached, although this is rather difficult to
determine because of the problems mentioned earlier. In every spectrum,
an absorption component is fitted at a velocity close to $-$2100
\kms. Probably, this component is at the terminal velocity of the wind
and the new DACs merge with this component when they approach their
asymptotic velocity (i.e.\ \vinf). Due to the above mentioned
limitations we cannot demonstrate this explicitly.

Signs of DACs are found in quotient spectra of the subordinate \heii\
(1640~\AA) and \niv\ (1718~\AA) lines (Fig.~27). The first traces of
these DACs occur at very low velocity: $-$200 \kms. A clear difference
with the behaviour of the \niv\ line in \xip\ is that the variations
in the \heii\ and \niv\ lines of \lab\ show the rapid acceleration of
DACs such as observed in, e.g., the \siiv\ resonance doublet. The
variations due to DACs in the \niv\ line of the O4 supergiant
$\zeta$~Pup are very similar (Prinja et al. \cite{PB92}). We conclude
that also in \lab\ the observations indicate that DACs are formed
already close to the star. Simultaneous H$\alpha$ observations support
this conclusion (Kaper et al. \cite{KH97}). 

The results of the Fourier analysis are presented in Fig.~28. The
summed 1d power spectra obtained for the October 1989 and February
1991 campaigns are shown for comparison. The October 1991 dataset
shows a period of 1.4$\pm$0.2~d (the solid line gives the power in
the red absorption component of \siiv\ between $-$600 and $-$1600
\kms); this period is also observed in February 1991, but then the
dominant period is [4.3$\pm$1.2~d]. Visual comparison of the time
series (Paper~I) suggests that the variability timescale in the
February 1991 data is indeed longer than in October 1991 (but the
sampling of the dataset is worse). The \niv\ profile
(long-dashed-dotted lines) produces the 1.4~d period (1.3$\pm$0.1~d)
too, the \heii\ line vaguely indicates a possible harmonic at 
0.72$\pm$0.03~d.  The October 1991 data also show evidence for a period
close to [5] days (although this is longer than the 4.5~d coverage). The
longer periods are predominantly found in the high-velocity edges of
the \siiv\ (long-dashed lines, $-$2200 to $-$2650 \kms) and the \civ\
(short-dashed lines, $-$2200 to $-$2750 \kms) P~Cygni profiles.

Although the October 1989 campaign lasted only 2.6 days, it appears
that the signature of the variability is different compared to October
1991 but rather similar to that in the February 1991 dataset. The
dominant period is at [2.5$\pm$0.8~d] in \siiv\ and [2.2$\pm$0.7~d] in
the \civ\ edge. Henrichs (\cite{He91}) showed that the optical \heii\
4686~\AA\ line exhibits (wind) variations that are in phase with the
modulation of the \civ\ edge, providing complementary support for the
significance of this 2-day period.

How can we reconcile the observed periods with each other? If 1.4~d
were the DAC recurrence time, then 2.8~d is two-times and 4.2~d
three-times this period. These periods are all within the error bars of
the observed ones. We will come back to this point in section 4.6.

The phase diagrams (Fig.~29) are difficult to interpret. The phase (in
radians, for power $>$5$\times$10$^{-5}$) related to the 1.4-day period
in October 1991 shows a decline in the red doublet component of the
\siiv\ doublet consistent with the DAC acceleration. In February 1991
the phase behaviour is not clear and in 1989 the 1.4-day period is not
detected (except in the blue edge). 
There is no clear evidence for phase bowing in \lab.

\subsection{HD~214680 (\lac) O9 V}

We modeled the DACs in quotient spectra of the O9~V star \lac\ which
was observed in November 1992 ($\triangle t =$ 5.0~d). DACs are present
in both the \nv\ and the \civ\ resonance doublet; here we show the
results from the profile fitting of the \nv\ doublet only, because the
DACs are most prominent in this line. The strongest component
(Fig.~30) appears at day 7.3 and is preceded by a DAC starting 1.4 day
earlier, which is not detected in the \civ\ doublet. The maximum
central velocity observed for both components is $-$990 \kms. Maximum
column density of the strongest component ($N_{\mbox{\scriptsize
col}}^{\mbox{\scriptsize max}} =$ 10$^{14}$ cm$^{-2}$) is reached when
$v_{c}$ is about $-$875 \kms. The time series is too short to
determine the recurrence timescale with confidence. The characteristic
timescale of variability is, however, longer than 4 days if the
modeled absorption components belong to the same event, as was
observed in the case of \xip\ and \cyg. Another argument in favour of
a timescale significantly longer than 1.4~day is that the acceleration
timescale, in this case at least several days, is usually comparable
to the DAC recurrence timescale.

The Fourier analysis applied to the \nv\ time series of \lac\ results
in a peak at [6.8$\pm$2.3~d] (and not at 1.4~d), but this period is
longer than the time coverage of the data. A more recent series of IUE
spectra, covering about 20 days, was obtained in August 1995 (Henrichs
et al., in prep.). In this series, the \nv\ and \civ\ doublet clearly
show two complete DAC events, with a recurrence timescale of about 7
days. A Fourier analysis yields a period of 6.8$\pm$1.0~d, 
consistent with the one detected in the much shorter November
1992 dataset.



\section{Comparing DAC behaviour in different stars}

Some of the general characteristics of the DAC phenomenon, such as the
narrowing of DACs when they accelerate through the profiles, their
always increasing speed until they reach an asymptotic velocity, and
their cyclical appearance are known for quite some time. In the
following, we will evaluate the DAC properties in O-star wind lines as
encountered in our observations.

\subsection{The DAC parameters} 

For the nine stars in our sample that show the presence and evolution
of DACs in one or more of the time series, we have quantitatively
modeled the DAC behaviour. We have demonstrated that the shape of the
individual absorption components can be very well described by an
exponential Gaussian. Table~2 summarizes the main results from the
fitting procedures. For each time series the extreme values of the fit
parameters are listed. Obviously, when more than one DAC event
was covered in the time series, the extremes do not necessarily
correspond to the same event. 

The central velocity asymptotically reached by a DAC at the end of its
track is a measure of the terminal velocity \vinf\ of the stellar wind
(Tab.~1). This velocity is typically 10--20\% less than the maximum
velocity in the steep blue edge of saturated P~Cygni lines, but about
equal to the velocity ($v_{\mbox{\scriptsize black}}$) corresponding
to their most blue-shifted point of saturation (Prinja et
al. \cite{PB90}). In radiation-driven winds \vinf\ is directly related
to the stellar escape velocity which does not change on short
timescales. In stars for which we collected more than one time series,
the measured maximum values of $v_{c}^{\mbox{\scriptsize asymp}}$ are
the same within 10\%. We note, however, that in some cases we found
systematic differences in $v_{c}^{\mbox{\scriptsize asymp}}$ from
event to event (see section 4.2). The data do not suggest a relation
between $v_{c}^{\mbox{\scriptsize asymp}}$ of a given DAC event and
its (maximum) strength ($\tau_{c}^{\mbox{\scriptsize max}}$ and/or
$N_{\mbox{\scriptsize col}}^{\mbox{\scriptsize max}}$).

\begin{figure*}[!ht]
\centerline{\psfig{figure=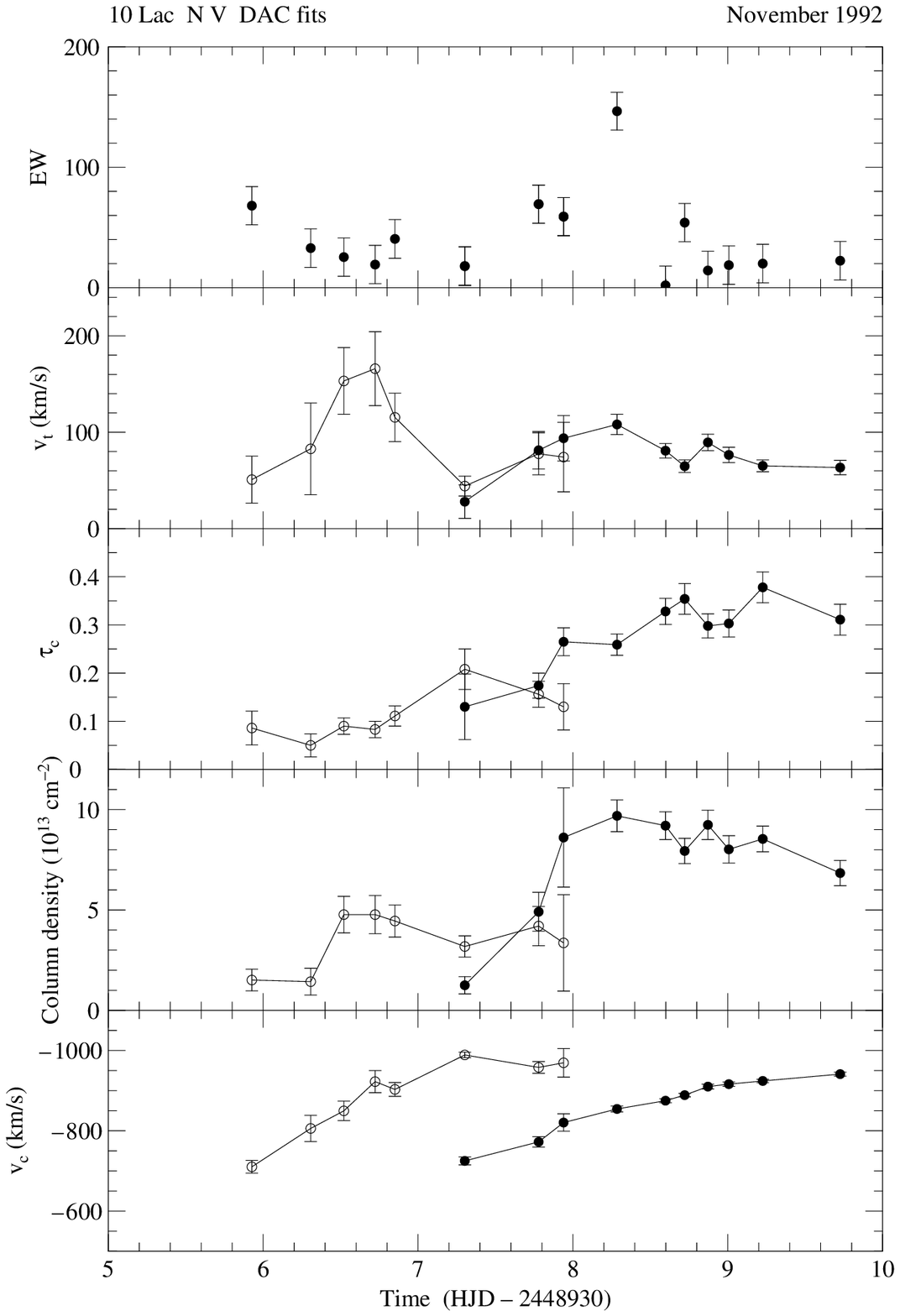,height=15cm}}
\caption[]{\lac\ O9~V November 1992 (as Fig.\ 7): DAC model parameters
for the \nv\ profile.}
\end{figure*}

The minimum central velocity, $v_{c}^{\mbox{\scriptsize min}}$, at
which DACs can be detected is typically in the range
0.2--0.4~\vinf. However, for the main-sequence stars in our sample
(\lor, \mon, \lac) $v_{c}^{{\rm min}}$ is of the order
0.7--0.9~\vinf. Also $\zeta$~Oph (O9.5~V, $v_{c}^{\mbox{\scriptsize
min}}\sim$1100~\kms, \vinf = 1480~\kms, Howarth et al. \cite{HB93})
with 0.74~\vinf\ and $\zeta$~Pup (O4~I(n)f, 
$v_{c}^{\mbox{\scriptsize min}}\sim$600~\kms, \vinf = 2520~\kms, 
Prinja et al. \cite{PB92}) with 0.24~\vinf\
fit this picture. There is no obvious link between the projected
rotational velocity $v \sin{i}$ and the lowest velocity at which DACs
are observed.  Given the large initial width of a DAC (sometimes up to
0.5~\vinf), though usually small initial optical depth, wind
variability can be traced down to zero velocity in several cases (see
also Paper~I). The difference in $v_{c}^{\mbox{\scriptsize min}}$
between main-sequence and evolved O~stars might be due to the
relatively weak winds of the main-sequence stars. Also the H$\alpha$
line of main-sequence stars does in general not show any signs of wind
variability (Kaper et al. \cite{KF98}).

\begin{table*}
\caption[]{Overview of DAC parameters of the studied O~stars after
modeling the resonance lines of \siiv\ or \nv (indicated with
$\star$, or both). The dominant period produced by the Fourier
analysis is called $P_{\mbox{\scriptsize wind}}$. Periods comparable
to or longer than the timespan $\triangle t$ of the dataset are put
between square brackets. The listed
values are yearly maxima and minima which do not necessarily belong to
the same DAC event.}
\begin{flushleft}
\begin{tabular}{lrlllllllll}
\hline\noalign{\smallskip}
Name & Yr & $v_{c}^{\mbox{\scriptsize min}}$ & $v_{c}^{\mbox{\scriptsize asymp}}$ & 
$\tau_{c}^{\mbox{\scriptsize max}}$ & $v_{t}^{\mbox{\scriptsize min}}$ & 
$v_{t}^{\mbox{\scriptsize max}}$ & $N_{\mbox{\scriptsize
col}}^{\mbox{\scriptsize max}}$ &
$v_{N{\mbox{\scriptsize max}}}$ & $P_{\mbox{\scriptsize wind}}$ & $\Delta t$ \\
     & & (\kms) & (\kms) & & (\kms) & (\kms) & (10$^{13}$cm$^{-2}$) &
(\kms) & (d) & (d) \\ \hline
\noalign{\smallskip}
\hline\noalign{\smallskip}
$\xi$ Per       & 87 & 1045$\pm$53 & 2438$\pm$8 & 1.71$\pm$0.04 
& \032$\pm$6 & 1221$\pm$110 & 49.3$\pm$1.1 & 1617 & 1.9$\pm$0.3 & 3.6 \\
                & 88 & \0400$\pm$141 & 2439$\pm$13 & 1.09$\pm$0.07 
& \054$\pm$7 & 2039$\pm$202 & 36.1$\pm$0.8 & 1499 & 1.0$\pm$0.1 & 3.0 \\
                & 89 & 1110$\pm$71 & 2141$\pm$23 & 1.05$\pm$0.04 
& \056$\pm$4 & \0691$\pm$13 & 43.0$\pm$0.7 & 1514 & 1.0$\pm$0.1 & 2.5 \\
                & 91 & \0660$\pm$73 & 2319$\pm$16 & 1.23$\pm$0.05 
& \039$\pm$11 & 1265$\pm$179 & 55.4$\pm$0.8 & 1771 & $2.0\pm$0.2 & 4.4 \\ \hline 
HD 34656        & 91 & \0548$\pm $25 & 2124$\pm$16 & 0.71$\pm$0.08
 & 112$\pm$16 & \0417$\pm$12 & 26.0$\pm$0.8 & 1606 &  1.1$\pm$0.1 & 5.1 \\
slow pattern      &       & 1300 & 1600 & 1.07$\pm$0.03 & \052$\pm$20
& \0436$\pm$12 & 24.4$\pm$0.6 & 1522 & [4.5$\pm$1.6] & \\ \hline
$\lambda$ Ori A & 92 & 1706$\pm$21 & 2000$\pm$15 & 0.31$\pm$0.03
 & \075$\pm$21 & \0236$\pm$20 & \04.5$\pm$0.3 & 1828 & [$\sim$4]? & 5.0 \\ \hline 
$\zeta$ Ori A   & 92 & \0522$\pm$19 & 2180$\pm$27 & 0.41$\pm$0.03 
& \046$\pm$12 & \0671$\pm$62 & \08.4$\pm$0.6 & 1129 & [6.3$\pm$2.1] & 5.1 \\
\nv\            &       & \0551$\pm$74 & 1983$\pm$31 & 0.21$\pm$0.02
 & \031$\pm$36 & \0454$\pm$78 & 19.8$\pm$2.6 & 1713 & [6.1$\pm$2.7] & \\ \hline
15 Mon$\star$   & 91 & 1711$\pm$38 & 2268$\pm$65 & 0.30$\pm$0.08 
& \043$\pm$16 & \0173$\pm$56 & \06.7$\pm$1.9 & 1711 & [5.8$\pm$1.8] & 5.3 \\ \hline
68 Cyg          & 86 & 1141$\pm$56 & 2499$\pm$5 & 1.09$\pm$0.03
& \035$\pm$8 & \0562$\pm$21 & 32.8$\pm$0.7 & 1587 & 1.4$\pm$0.1& 6.6 \\
                & 87 & 1200$\pm$40 & 2448$\pm$5 & 0.82$\pm$0.03
& \097$\pm$15 & \0701$\pm$24 & 25.9$\pm$0.7 & 1641 & 1.3$\pm$0.2 & 3.4 \\
                & 88 & \0588$\pm$332 & 2565$\pm$13 & 1.03$\pm$0.04 
& \026$\pm$9 & 1464$\pm$414 & 28.9$\pm$1.2 & 1627 & 1.3$\pm$0.2 & 3.1 \\
                & 89 & \0814$\pm$41 & 2419$\pm$11 & 1.00$\pm$0.10 
& 118$\pm$20 & \0998$\pm$64 & 23.8$\pm$0.5 & 2187 & [3.0$\pm$0.9] & 2.6 \\
                & 91 & \0840$ \pm$27 & 2488$\pm$7 & 0.88$\pm$0.03 
& \030$\pm$10 & 1258$\pm$111 & 28.3$\pm$0.6 & 1916 & 1.4$\pm$0.2 & 4.5 \\ \hline
19 Cep          & 86 & \0317$\pm$94 & 1936$\pm$17 & 2.28$\pm$0.05 
& \064$\pm$18 & 1010$\pm$104 & 68.3$\pm$1.1 & 1455 & [7.0$\pm$2.3] & 6.7 \\
\nv\            &       & \0545$\pm$26 & 1919$\pm$26 & 1.13$\pm$0.05 
& \042$\pm$18 & \0412$\pm$100 & 118.1$\pm$4.7 & 1500 & [7.0$\pm$2.3] & \\ 
                & 91 & \0454$\pm$63 & 2087$\pm$23 & 0.78$\pm$0.04 
& \039$\pm$7 & \0846$\pm$120 & 29.1$\pm$1.0 & 1494 & [6.3$\pm$2.1] & 4.2 \\ 
\nv\            &       & 1137$\pm$181 & 2067$\pm$18 & 0.80$\pm$0.10 
& \027$\pm$19 & 1185$\pm$222 & 67.6$\pm$5.2 & 1829 & [5.6$\pm$1.9] & \\ 
                & 92 & \0771$\pm$17 & 1815$\pm$12 & 1.14$\pm$0.04 
& \099$\pm$7 & \0459$\pm$19 & 34.1$\pm$0.8 & 1561 & [4.5$\pm$1.9] & 5.6 \\ 
\nv\            &       & \0559$\pm$35 & 1864$\pm$19 & 0.90$\pm$0.05 
& 125$\pm$19 & \0450$\pm$69 & 101.0$\pm$4.7 & 1647 & [5.1$\pm$1.3] & \\ \hline 
$\lambda$ Cep   & F91 & \0703$\pm$21 &              & 1.57$\pm$0.07 
&             & \0484$\pm$36 &              &      &[4.3$\pm$1.2] & 5.1 \\ 
                & O91 & \0481$\pm$22 &              & 1.26$\pm$0.10 
&             & \0470$\pm$51 &              &      & 1.4$\pm$0.2 & 4.5 \\ \hline
10 Lac$\star$   & 92 & \0710$\pm$16 & \0989$\pm$7 & 0.38$\pm$0.03 
& \044$\pm$10 & \0166$\pm$38 & \09.7$\pm$0.8 & \0854 & [6.8$\pm$2.3] & 5.0 \\ \hline
\noalign{\smallskip}
\hline
\end{tabular}
\end{flushleft}
\end{table*}

The maximum optical depth ($\tau_{c}^{\mbox{\scriptsize max}}$)
encountered for these absorption features is around one. The strongest
DACs are found in the supergiant and giant spectra, although \zor\ has
relatively weak absorption components for a supergiant and \xip\
relatively strong components for a giant. The DACs in main-sequence
stars have the smallest optical depths. The minimum DAC widths
($v_{t}^{\mbox{\scriptsize min}}$) we measured are several tens to a
hundred \kms. The resolution of the SWP camera aboard IUE is about
25~\kms.

The DAC maximum column density varies from event to event, but in
general $N_{\mbox{\scriptsize col}}^{\mbox{\scriptsize max}}$
(i.e. the highest value of $N_{\mbox{\scriptsize col}}$ in a given
time series, often containing several DAC events) is largest for the
supergiants and smallest for the main-sequence stars in our
sample. The velocity at which a maximum in column density is reached
is always in the range 0.7--0.9~\vinf, i.e.  about half-way the
evolution of a DAC (Figs. 31 and 32).  For \zor\ and \cep\ the DACs
were fitted in both the \siiv\ and the \nv\ doublet. The results are
consistent; the measured $N_{\mbox{\scriptsize col}}$ is about a
factor 2 higher in the \nv\ doublet. With reference to Eq.~10, the
difference in oscillator strength ($f(\mbox{\siiv})/f(\mbox{\nv})=$
3.4) is the main factor.

\begin{figure}[!ht]
\centerline{\psfig{figure=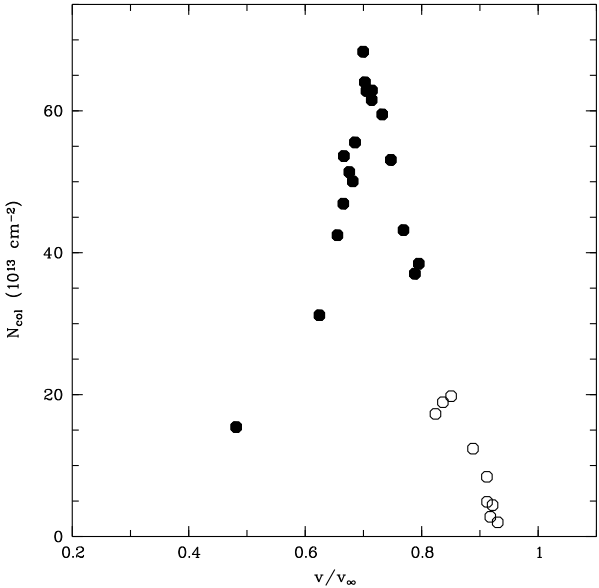,width=8cm,angle=-90}}
\caption[]{The variation of DAC column density as a function of
central velocity in the \siiv\ resonance doublet of \cep. The filled
diamonds represent the strong DAC event in the August 1986 dataset;
the open circles are the corresponding values for the DAC already at
$v_{c}^{\mbox{\scriptsize asymp}}$ at the beginning of the 1986
campaign. Clearly, the DAC reaches a maximum in $N_{\mbox{\scriptsize
col}}$ at 0.7~$v_{\infty}$.}
\end{figure}

\begin{figure}[!t]
\centerline{\psfig{figure=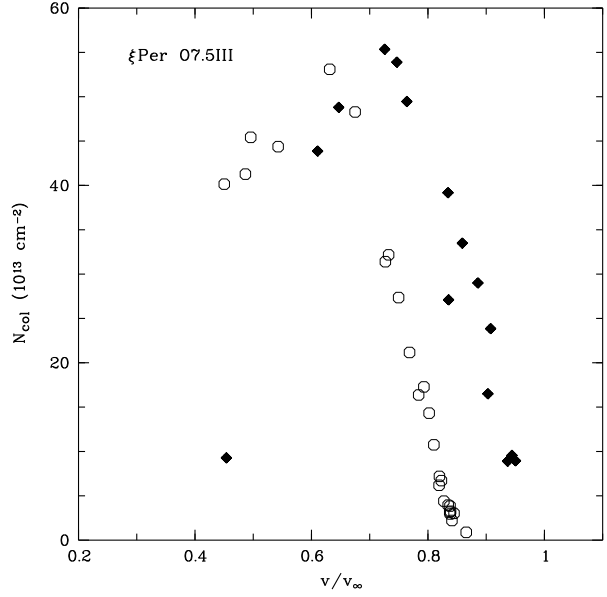,width=8cm,angle=-90}}
\caption[]{The variation of DAC column density as a function of
central velocity in the \siiv\ resonance doublet of \xip\ (October
1991). The open circles represent the ``first'' DAC event that reaches
a lower $v_{c}^{\mbox{\scriptsize asymp}}$ than its successor (filled
diamonds). Although the two curves reach the same maximum in
$N_{\mbox{\scriptsize col}}$, they are shifted in velocity.}
\end{figure}

\subsection{Short-term DAC evolution}

The most striking property of the DAC phenomenon is the regularity in
recurrence over many years. Time-series analyses show that the
cyclical behaviour of DACs often results in a well-determined ``wind''
period. Although also other periods are found (some of them can be
identified as harmonics of the principal period), the period with
strongest power ($P_{\mbox{\scriptsize wind}}$) most often corresponds
to the DAC recurrence timescale $t_{\mbox{\scriptsize rec}}$. In the
case of \zor\ $t_{\mbox{\scriptsize rec}}$ (1.6$\pm$0.2~day) is also
detected in the power spectrum, but most power is associated with a
period of about [6] days. In Table~2 we list $P_{\mbox{\scriptsize
wind}}$ derived from the \siiv\ (or \nv) resonance lines. The shorter
the time span $\Delta t$ of the campaign, the less reliable the
detected periods. When the period is comparable to or longer than 
$\Delta t$ it cannot be trusted. In Table~2 we marked these periods by
putting them in between square brackets.

The DAC recurrence timescale is shorter for stars with a higher
projected rotational velocity (Henrichs et al. \cite{HK88}, Prinja
\cite{Pr88}); this conclusion is supported by the results based on our
larger sample. In Table~3 we compare the recurrence timescale of DACs
with the estimated maximum rotation period of the star (the
inclination of the stellar rotation axis is not known). An estimate
for the stellar radius is obtained from Howarth \& Prinja
(\cite{HP89}). For all stars $t_{\mbox{\scriptsize rec}}$ is smaller
than the maximum rotation period of the star. This supports the view
that stellar rotation determines the characteristic timescale of
(large-scale) wind variability in O-type stars. Whether
$t_{\mbox{\scriptsize rec}}$ is equal to the stellar rotation period
(or an integer fraction of this) will be discussed in
section~5. Coordinated H$\alpha$ and UV observations of a few bright
O~stars have shown that $P_{\mbox{\scriptsize wind}}$ detected in UV
resonance lines also appears in the H$\alpha$ line (Kaper et
al. \cite{KH97}). On the basis of a large H$\alpha$ survey Kaper et
al. (\cite{KF98}) demonstrate that the relation between
$P_{\mbox{\scriptsize wind}}$ and the maximum rotation period also
holds for this larger sample.

The variability of the inner regions of the stellar wind, where
H$\alpha$ is formed, is directly linked to the DAC behaviour in UV
resonance lines. This is confirmed by the detection of similar
variability in subordinate UV lines in \xip, \sao, and \lab.  In these
stars, the subordinate \niv\ line at 1718~\AA\ with P~Cygni-type
profile (and for \lab\ also the \heii\ ``Balmer$\alpha$'' at 1640~\AA)
varies at low velocity in concert with the DACs (see also Henrichs et
al. \cite{HK94} and Paper~I). DACs have been detected in the \niv\
line of $\zeta$~Pup too (Prinja et al. \cite{PB92}). As these lines
have to be formed in regions of higher density, this implies that
material forming DACs must be present already close to the star
(within a few $R_{\star}$).

Not only the cyclical appearance, but also the change in DAC central
velocity per unit time is larger for stars with a shorter rotation
period.  This becomes apparent when one compares the DAC behaviour of
the rapid rotators (e.g.\ \xip, \lab) with that in the most-likely
slowly rotating stars (e.g.\ \cep, \lac). This suggests that the
``acceleration'' of DACs is not only a consequence of the line-forming
region moving away from the star (where the wind speeds are higher);
stellar rotation must play an important role too in explaining the
acceleration timescale.

The variation of DAC column density as a function of their central
velocity (i.e. time) is nicely demonstrated by the apparently slow
rotator \cep\ and the rapid rotator \xip\ in Figs.\ 31 and 32,
respectively. In both cases, the dependence of $N_{\mbox{\scriptsize
col}}$ on $v_{c}$ resembles a parabolic function; a maximum in
$N_{\mbox{\scriptsize col}}$ is reached half way. As \vinf\ we took
the maximum $v_{c}^{\mbox{\scriptsize asymp}}$ we observed for the
star (Table~1). It turns out that, for a given star, not all DACs have
the same $v_{c}^{\mbox{\scriptsize asymp}}$. This can vary from
campaign to campaign, or from event to event (e.g. \xip,
Fig. 32). Although the absolute velocities are different, the overall
shape of the variation in $N_{\mbox{\scriptsize col}}$ is very
similar, even for different stars.

As said before, the strength of a DAC varies from event to
event. Sometimes, a weak absorption component disappears before
reaching the wind terminal velocity. In several stars (e.g. \xip\ and
\cyg) the stronger DACs are accompanied by weak absorption components
separated by about a quarter of the DAC recurrence timescale. In most
cases the two components merge when they arrive at the same velocity.

\begin{table}
\caption[]{DAC recurrence timescale compared to the estimated maximum
stellar rotation period $P_{\mbox{\scriptsize max}} =$ 50.6 $(v
\sin{i})^{-1} (R_{\star}/R_{\odot})$~days for the program stars. An
estimate for the stellar radius is taken from Howarth \& Prinja
(\cite{HP89}). The stars are ordered according to $P_{\mbox{\rm
max}}$. The last column lists the proposed stellar rotation period,
based on our interpretation of the DAC behaviour.}
\begin{flushleft}
\begin{tabular}{lccccc}
\hline\noalign{\smallskip}
Name & $v \sin{i}$ & $R_{\star}$ & $t_{\mbox{\scriptsize rec}}$ & 
$P_{\mbox{\scriptsize max}}$ & $P_{\mbox{\scriptsize rot}}$ \\
     & (\kms) & ($R_{\odot}$ & (d)  & (d) & (d) \\ \hline
\noalign{\smallskip}
\hline\noalign{\smallskip}
68 Cyg          & 295 & 14 & 1.4 & \02.4 & 2.8 \\
$\xi$ Per       & 204 & 11 & 2.0 & \02.7 & 4.0 \\
$\lambda$ Cep   & 214 & 19 & 1.4 & \04.0 & 2.8 \\
HD 34656        & \085 & 10 & 1.1 & \06.0 & 4.5 \\
15 Mon          & \062 & 10 & $>$4.5 & \08.2 & $>$5 \\
$\lambda$ Ori A & \066 & 12 & $\sim$4? & \09.2 & $>$4 \\
19 Cep          & \090 & 18 & $\sim$ 5 & 10.1 & 5 or 10 \\
$\zeta$ Ori A   & 123 & 31 & 1.6 & 12.8  & 6? \\
10 Lac          & \031 & \09 & 6.8 & 14.7 & 7 or 14 \\ \hline
\noalign{\smallskip}
\hline
\end{tabular}
\end{flushleft}
\end{table}

In \xip\ and \cyg\ subsequent (strong) DACs alternate in
asymptotic velocity. For \xip\ the strong components reach $-$2050 or
$-$2300 \kms. Since the lifetime of a strong DAC is longer than two
days (the recurrence timescale), this results in ``crossings'' of
successive components; they arrive at the same velocity, which does
not mean that the material forming the two DACs is at the same
physical location (though both in the line of sight). For \cyg\ we
identified two series of DACs, series~A reaching an asymptotical
velocity of $-$2450~\kms\ and series~B $-$2200 \kms. A difference with
\xip\ is that the time interval between series~B and A is 0.5~d, which
is much shorter than the recurrence timescale of 1.4~d (the distance in
time between subsequent A's or B's, as well as pairs B~A). 

\noindent
The difference between these sequences of DACs is visualized as 
follows:\\[0.2cm]
\begin{tabular}{|lcccccccccc|}\hline
     & \multicolumn{5}{l}{series A $\rightarrow v_{\rm asymp}^{1}$} &
\multicolumn{5}{l|}{series B $\rightarrow v_{\rm asymp}^{2}$} \\ \hline
\xip & A & B & A & B & A & B & A & B & A & \dots \\ \hline
\cyg & A &   & B & A &   & B & A &   & B & \dots \\ \hline
\sao & A & B & B & B & A & B & B & B & A & \dots \\ \hline
     & \multicolumn{10}{l|}{time $\rightarrow$} \\ \hline
\end{tabular}
\vspace{0.2cm}

The above mentioned DAC properties make up for the observed
``pattern'' of variability that is characteristic for the
star. Sometimes it seems to be more complicated; e.g. in \sao\ two
different patterns can be recognized, a fast one and a slow one.  More
extended time series, like those obtained during the IUE MEGA campaign
(Massa et al. \cite{MF95}), also indicate that the regular variability
pattern is modulated by variations occuring over a longer, but
eventually related timescale. A good example is provided by the MEGA
campaign data on $\zeta$~Pup (Howarth et al. \cite{HP95}). These
authors interpret the longer (5.2-day) period as the stellar rotation
period; the shorter (19-hour) recurrence timescale of the DACs would,
however, not be an integer fraction (6.5$\pm$0.9) of the rotation
period. If the latter were true for this star, stellar rotation alone
would not be sufficient to explain the cyclical recurrence of DACs.

\subsection{Edge variability}

In saturated P~Cygni profiles wind variability manifests itself in
significant shifts in velocity of the steep blue edge (cf.\
Paper~I). In section~3 it is shown that close to the high-velocity
edges of the UV resonance lines often a longer period is detected than
the period exhibited by the DACs. We propose that the edge variability
is directly related to the DAC behaviour. The observations suggest
that the slow modulation of the high-velocity edges is the result of
the fact that only the DACs with the highest asymptotic velocity have
an impact on the position of the edge. The case of \xip\ provides the
strongest support to this interpretation: the edge variability has a
period of about 4 days (i.e., twice the DAC recurrence timescale),
while only one out of two DACs reaches an asymptotic velocity
exceeding $-$2050 \kms. 

The combination of a slow and a more rapid variability pattern (such
as observed in \sao and $\zeta$~Pup) and the subsequent modulation in
strength of the DACs might give rise to edge variability in saturated
profiles on the timescale of the slow pattern. Therefore, only
extended time series (such as those obtained in the IUE MEGA campaign,
cf.\ Massa et al. \cite{MF95}) might enable confirmation of this
conjecture.

\subsection{Phase diagrams}

The phase diagrams included in this paper show the variation of the
sinusoid's phase at the dominant wind variability period with position
in the line. The observed decrease in phase towards higher velocities
is due to the acceleration of the DACs through the line profile. This
supports our view that $P_{\mbox{\scriptsize wind}}$ is equal to the DAC
recurrence timescale. The large width of the DACs when they appear at
low or intermediate velocity would explain the relatively small change
in phase observed at these velocities.

Owocki et al. (\cite{OC95}) discuss the phase ``bowing'' discovered in
the phase diagrams belonging to $P_{\mbox{\scriptsize wind}}$ detected
in the B0.5~Ib star HD~64760 (Prinja et al. \cite{PM95}, Fullerton et
al. \cite{FM97}). Although, according to these authors, in this star
$P_{\mbox{\scriptsize wind}}$ (1.2 or 2.4~d) is caused by a modulation
in flux at low and intermediate velocities rather than by DACs, the
decrease of phase towards both higher and lower velocities (which
appears as a bow in the phase diagram) is a strong diagnostic of
curved wind structures (like co-rotating interacting regions,
section~5). DACs are detected at higher velocities in the resonance
lines of HD~64760 and are variable on a timescale longer than
$P_{\mbox{\scriptsize wind}}$.

The phase diagrams for the 2-day period detected in the \xip\ October
1991 dataset clearly show evidence for phase bowing. A maximum in
phase occurs around $-$1200		~\kms, i.e. about the central velocity at
which the DACs usually appear (Fig.~8). The maximum phase lag $\triangle
\phi$ is about one radian, similar to the phase lag observed in
HD~64760. The increasing strength of the DAC might explain the
subsequent decrease in phase towards higher and lower velocities. The
further acceleration and narrowing of the DACs is reflected by the
increasing slope of the phase towards higher velocities. This
description of the phase behaviour is only qualitative; detailed
modeling of the phase diagram falls beyond the scope of this paper. An
important difference with HD~64760 is that the DACs themselves seem to
be responsible for the observed phase bowing.

Phase bowing is not evident in the phase diagrams for the other
O~stars in our sample. It has only been found in \xip, which is
perhaps not surprising, because the other stars in our sample do not
exhibit such strong wind variability down to the lowest wind
velocities.

\subsection{Flux modulation and DAC behaviour}

The ``classical'' picture we have of DAC behaviour, and which is
supported by the observations analysed in this paper, is that broad
absorption components appear at low velocity which evolve towards
higher velocity while narrowing. The application of Fourier analyses
resulted in the determination of well-constrained periods, not only at
low velocity, but over the full width of the lines. To check the
outcome of the Fourier analysis we constructed diagrams showing the
variation of flux as a function of time in a each velocity bin. It
indeed appears that the flux varies sinusoidally with time, especially
at low and intermediate velocities. In some cases the Fourier analysis
invokes an additional sinusoid to accommodate the weak precursor of the
next DAC (resulting in the addition of a shorter-period sinusoid) or a
gradual change (longer period). The latter occurs for example when two
subsequent DACs reach a different asymptotic velocity.

The analysis of two spectroscopic time series of UV resonance lines of
the B0.5~Ib supergiant HD~64760 lead Fullerton et al. (\cite{FM97}) to
conclude that in this star a distinction can be made between a
sinusoidal modulation in flux at low and intermediate wind velocities
and variations due to the propagation of DACs at higher velocities. A
2.4-day (or 1.2-day) periodicity is connected with the modulation
which clearly shows the occurrence of phase bowing; the DACs vary on a
longer timescale. Thus, in HD~64760 two different kinds of wind
variability are observed: flux modulation and DAC propagation. Because
of the observed phase bowing, Fullerton et al. (\cite{FM97}) interpret
the modulation to be due to CIRs, whereas the origin of the DACs is
not clear.

Our results on \xip\ indicate a different picture: in this star the
flux modulation and phase bowing are caused by the DACs themselves.
That the propagation of DACs leads to a sinusoidal modulation of the
flux in a given velocity bin is due to the shape of the DACs, which
can be accurately fit with an exponential Gaussian. At low velocity
these gaussians have widths of several hundreds km/s. At high velocity
they become very narrow. When the DAC moves through the resonance
line, the flux in a given velocity bin changes gradually, following
the Gaussian shape of the DAC. At high velocities, where the DACs are
narrow, the flux changes more irregularly, because the DACs ``pile
up'' at about the same asymptotic velocity before they vanish.

This might explain why the flux gradually changes when a DAC moves
through a velocity bin, but not why this modulation is sinusoidal. In
our interpretation, this is because not only the recurrence timescale,
but also the acceleration timescale of the DAC is proportional to the
rotation period of the star. Since these two timescales are identical,
a sinusoidal shape of the flux versus time plots is expected. Thus,
the DAC behaviour in O-star winds can result in a sinusoidal
modulation of the flux in each velocity bin.

With our definition of the template, the maximum of the sinusoidal
modulation corresponds to the ``undisturbed, underlying wind
profile'', so that all variations are by definition due to additional
absorption. From the point of view of Fullerton et al. (\cite{FM97}),
in the case of HD~64760 the template might be better defined as the
zero-point of the modulation, i.e. the ``average wind profile''. With
such a template, it is not possible to model the DACs they observe in
the resonance lines of HD~64760 at high velocity as absorption
components, which is the reason for our approach.

\subsection{Long-term variability}

The O~stars \xip, \cyg, \cep, and \lab\ were observed several times in
the period 1986--1992. The variability pattern, characteristic for a
given star, can be recognized from year to year, though detailed
changes occur. The same wind variability timescale
$P_{\mbox{\scriptsize wind}}$ is detected every year. Sometimes a
different period gives the highest peak in the power spectrum (\xip\
1988 and 1989, \cyg\ 1989), but this might be due to the short time
coverage of these datasets. For \lab\ there seems to be a clear
difference between the observed variability timescale in October 1989,
February 1991, and October 1991. In 1989 a period of [2.5$\pm$0.8~d] is
dominant, in February 1991 a period of [4.3$\pm$1.2~d] (as well as 
1.4$\pm$0.1~d), and in October 1991 a period of 1.4$\pm$0.2~d. If stellar
rotation is setting the timescale of wind variability, one would not
expect this timescale to vary from year to year. What could in
principle vary is the number of features observed per rotation
period. In the case of \lab\ the difference in observed periods could
be explained as different integer fractions of the stellar rotation
period: 1.4, 2.8, 4.2~d; or 1.3, 2.6, 3.9~d. In section~5 we will
address the point whether the DAC recurrence timescale can be used to
determine the stellar rotation period.

Although the characteristic variability timescale remains the same,
the DAC strengths and asymptotic velocities exhibit changes from
campaign to campaign. The DAC behaviour in \cyg\ seems to be the most
constant over the years; clear differences are observed for \xip\
(Fig.~6), \cep, and \lab. This indicates that the underlying clock
might be the same, but that at least for some stars in our sample the
pattern of large-scale structure in the stellar wind varies on a
timescale of months to years.

\section{Towards an interpretation}

Several models have been put forward to explain the observed
properties of DACs. The expansion of a high-density layer in the
stellar wind (Henrichs et al. \cite{HH83}) gives a simple explanation
for the time-dependent behaviour of DACs, but it gives no clue to the
origin of the high-density layer. Underhill \& Fahey (\cite{UF84})
tried to explain the existence and behaviour of DACs by postulating
the existence of ``closed'' and ``open'' magnetic loops above the
stellar surface from which parcels of gas are released in addition to
a uniformly emitted steady stellar wind. A nice property of
this model is that DACs are always expected to appear at the same
rotational phase as long as the magnetic footpoints remain at the same
location. 

DACs are found at the same velocity in resonance lines formed by
different ions (e.g.\ Lamers et al. \cite{LG82}); therefore, DACs
most-likely originate from regions which have a higher density (or
smaller velocity gradient) than the ambient wind. These regions have
to be geometrically extended, because: (1) practically all O~stars
show DACs; (2) they have to cover a significant fraction of the
stellar disk to be observable.

Stellar rotation plays a dominant role in setting the timescale of
variability; a difference in flow time scale ($R_{\star}/v_{\infty}$)
cannot explain the observed dependence on $v \sin{i}$. Not only the
cyclical recurrence, but also the acceleration of DACs is faster for
more rapidly rotating stars. This suggests that the DAC forming region
rotates into and out of the line of sight. The observed increase and
subsequent decrease of $N_{\mbox{\scriptsize col}}$ supports this
picture (Figs. 31 and 32).

\subsection{Corotating Interacting Regions and DACs}

Most-promising is the Corotating Interacting Regions model proposed by
Mullan (\cite{Mu84}, \cite{Mu86}), in analogy with the corotating
interacting regions in the solar wind which result from the
interaction of fast and slow wind streams. Recently, this model has
been worked out by Cranmer \& Owocki (\cite{CO96}) who performed
two-dimensional hydrodynamical simulations of corotating stream
structure in the wind of a rotating O~star. The corresponding
spiral-shaped, large-scale structure through which the wind material
is flowing, was proposed by Prinja \& Howarth (\cite{PH88}) based on
observations of \cyg. Simultaneous ultraviolet and H$\alpha$
spectroscopy of a number of bright O~stars included in this study
strongly support the CIR model (Kaper et al. \cite{KH97}, see also
Fullerton et al. \cite{FM97}). Cranmer \& Owocki (\cite{CO96})
conclude from their model that the DACs are not formed in the CIRs
themselves, but originate in the so-called radiative-acoustic kinks
trailing the CIRs. The strong velocity gradient related to this kink
results in a relatively large contribution to the Sobolev optical
depth. Kaper et al. (\cite{KH97}) show that a small phase lag might be
present between the wind variations in the H$\alpha$ line with respect
to those observed in the UV resonance lines; this would be consistent
with the H$\alpha$ variations originating in the CIR and the DACs in
the trailing kink.

Phase bowing suggests that the structures are curved, explaining why
material at a given radial velocity rotates out of or into the line of
sight first, followed by material moving at higher and lower
velocities (Owocki et al. \cite{OC95}, Cranmer \& Owocki
\cite{CO96}). All nine O~stars with DACs show a decrease in phase
towards higher velocities; only one (\xip) also shows a lag in phase
towards lower velocities. The location $r_{i}$ in the wind where the
fast flow first meets the slow stream depends on the equatorial
rotation speed $v_{\mbox{\scriptsize rot}}$ and the difference in
(terminal) velocity between the fast and slow stream (Mullan
\cite{Mu84}):
\begin{equation}
\frac{r_{i}}{R_{\star}} = 1 + \frac{\triangle \psi}{v_{\mbox{\scriptsize
rot}}} \left[ \frac{v_{\infty}^{\mbox{\scriptsize fast}}
v_{\infty}^{\mbox{\scriptsize slow}}}{v_{\infty}^{\mbox{\scriptsize
fast}} - v_{\infty}^{\mbox{\scriptsize slow}}} \right]
\end{equation}
Inwards of $r_{i}$, the stellar wind might be rather smooth, so that
variability is not detected at low wind velocities. Also, the
``contrast'' between the fast and slow stream (angular separation
$\triangle \psi$ and $v_{\infty}^{\mbox{\scriptsize fast}} -
v_{\infty}^{\mbox{\scriptsize slow}}$) is likely to vary from star to
star. For large $r_{i}$ it might well be that only a decrease in phase
towards higher velocities will be detected. Furthermore, the aspect
angle might be an important factor; e.g. the decrease in phase towards
lower velocities is only observed when the line of sight includes the
equatorial plane.

The CIR model only works for rotating stars with emerging wind flows
of different kinematic structure. Comparison between model and
observations suggest that wind streams with relatively slow speed
interact with the faster ambient wind, in stead of high-speed streams
in a slow wind (Cranmer \& Owocki \cite{CO96}). Therefore, the
asymptotic velocity reached by DACs might be somewhat lower than the
terminal velocity of the ambient wind. In some stars we measure
systematically different values of $v_{c}^{\mbox{\scriptsize asymp}}$ from
event to event. This might be due to a difference in aspect angle,
contrast, or both.

\subsection{The origin of CIRs: non-radial pulsations or surface
magnetic fields?}

In order to work, this model needs a certain structure imposed at the
stellar surface to produce flows with different kinematical
properties. Two physical mechanisms that in principle could cause fast
and slow streams in the wind are: (i) non-radial pulsations (NRP), or
(ii) surface magnetic fields. Henrichs (\cite{He84}) pointed out that
the occurrence of DACs might be related to the presence of NRP at the
stellar surface. NRP are detected (or suspected) in several O~stars
(cf.\ Baade \cite{Ba88}, Fullerton et al. \cite{FG96}). The associated
temperature and velocity fields (on the order of the sound speed)
could be responsible for the modulation of the boundary conditions at
the base of the radiation-driven wind.

Our working hypothesis is that the outflow properties at the base of
the wind are influenced by a magnetic field anchored in the star,
which results in regions with different outflow properties rotating
with the star. The presence of magnetic fields in early-type stars is
difficult to prove with direct observational methods. Magnetic fields
with a strength of less than 100 Gauss are theoretically predicted
(Maheswaran \& Cassinelli \cite{MC92}), but not (yet) observable
(Landstreet \cite{La92}). Recently, Stahl et al. (\cite{SK96}) argued
that a surface magnetic field might be present in the O7~V star
$\theta^{1}$~Ori~C; phase-locked photospheric and stellar-wind
variations strongly suggest that this star is an oblique magnetic
rotator. An upper limit of 70~G for the (longitudinal) magnetic field
strength of \xip\ has been derived by Henrichs et
al. (\cite{HD98}). Recent attempts to measure the longitudinal
magnetic field strength in a sample of 10 fairly slowly rotating
O~stars by Mathys (\cite{Ma98}) and collaborators did not result in a
definite detection (1$\sigma$ error bars of the order of
250--300~G). Some marginal measurements (between 2 and 3$\sigma$)
suggest that weak fields might nevertheless be present, which could
become detectable through a fairly modest improvement of the
measurement accuracies.

Early-type stars are relatively short-lived objects and born out of
interstellar matter that contains a magnetic field. The star with the
strongest magnetic field known, ``Babcock's star'' HD 215441, is a
B3~V star with a surface dipole field strength of 34 kGauss (Babcock
\cite{Ba60}, Preston \cite{Pr71}). This demonstrates that early-type
stars can have very strong magnetic fields (see also Bohlender
\cite{Bo94}). These fields are expected to be frozen-in into the
radiative envelopes of the stars (see, for example Mestel
\cite{Me75}). Furthermore, the end-products of early-type stars are
mostly neutron stars with a strong magnetic field, which strength can
be predicted by assuming magnetic flux conservation during the
collapse of the core and the presence of a surface magnetic field of
strength 10--100~G in the progenitor. Alternative scenarios for the
formation of neutron-star magnetic fields exist as well.

\subsection{Can $t_{\mbox{\scriptsize rec}}$ be used to measure
$P_{\mbox{\scriptsize rot}}$?}

When we adopt the CIR model in combination with an anchored surface
magnetic field to explain the presence and cyclical behaviour of DACs,
the characteristic timescale of variability in the winds of O-type
stars {\it is} the rotation period of the star. Our observations
suggest that in several cases (e.g.\ \xip\ and \cyg\ where successive
DACs have different asymptotical velocities) the rotation period of
the star is more likely equal to twice (or an integral number times)
the recurrence timescale of DACs. This means that DACs are not only a
diagnostic for deriving the terminal velocity of a stellar wind,
but they can in principle be used to derive the rotation period
of the star. In Table~3 we list, based on the interpretation of the
cyclical DAC behaviour, the best estimate for the rotation period of
the star. Given an estimate of the stellar radius, the inclination of
the rotation axis can thus be derived from $v \sin{i}$.

In two cases (\xip\ and \cyg) the proposed rotation period exceeds the
estimated maximum rotation period. It is likely that these rapidly
rotating stars (on the basis of the measured $v\sin{i}$) have a
$\sin{i}$ value close to unity, and the radius of these stars could be
slightly underestimated. For \xip\ a radius of 15$R_{\odot}$ instead
of 11$R_{\odot}$ would increase the maximum rotation period to 4 days
(see Puls et al.\ \cite{PK96}). Walborn et al.\ (\cite{WN85}) note
that \xip\ is of luminosity class II instead of III which supports
this conclusion. The uncertainty in the spectral class of \cyg\
(denoted by the colon) also indicates that the estimated radius might
not be accurate.

\subsection{Summary and future work}

The DAC behaviour in the UV resonance lines of a sample of 10 bright
O~stars has been analysed in detail. The cyclical appearance and
subsequent evolution of DACs is interpreted in terms of a model
invoking fast and slow streams which interact due to the rotation of
the underlying star. The interacting regions (CIRs) corotate with the
star while the wind material is flowing through them. The observations
suggest that the stellar wind includes more than one CIR, most likely
two (cf.\ Kaper et al. 1997). The curvature of the CIRs can in
principle explain the observed phase behaviour (bowing), which is
detected in one of the O~stars in our sample. The wind variability
periods we derive are a direct measure of the DAC recurrence
timescales and, in our interpretation, equal to an integer fraction
(most likely 1/2) of the stellar rotation period.

The key issue is the origin of the CIRs: are non-radial pulsations,
surface magnetic fields, or other physical mechanisms responsible for
the surface structure creating fast and slow wind streams? This is
the subject of our current investigations.

\begin{acknowledgements}
The authors wish to thank NASA and ESA/SERC for the generous amount of
IUE observing time dedicated to this program. We are particularly 
grateful to the late Dr.\
Roelf Takens for bringing the paper of Peat \& Pemberton to our
attention and for his valuable suggestions. We thank the
referee Dr.\ Alex Fullerton for a thorough reading of the manuscript
and helpful remarks. Support of the Netherlands Organization for
Scientific Research (NWO) is acknowledged by LK under grant
782-371-037 and by JHT under grant 781-71-043.
\end{acknowledgements}

\end{document}